\documentclass[%
preprint, 
superscriptaddress,
 amsmath,amssymb,
 aps, 
]{revtex4-2}

\usepackage{graphicx}
\usepackage{dcolumn}
\usepackage{bm}

\usepackage{amsmath}
\usepackage{amssymb}
\usepackage{graphicx}
\usepackage{subcaption}
\usepackage{makecell}
\usepackage{blkarray}
\usepackage{mathtools}
\usepackage{xcolor}
\usepackage{soul}
\usepackage{hyperref}
\usepackage{mathrsfs}
\usepackage{xspace}
\numberwithin{equation}{section}

\newcommand{\dz}{\mathrm{d}z}

\newcommand{\dee}{\mathrm{d}}

\newcommand{\KD}{K\'{a}rm\'{a}n--Donnell }
\newcommand{\Pcal}{\mathcal{P}}
\newcommand{\Lcal}{\mathcal{L}}
\newcommand{\Scal}{\mathcal{S}}
\newcommand{\Tcal}{\mathcal{T}}

\renewcommand{\colorbox}[2]{#2}
\renewcommand{\hl}[1]{#1}

\begin{document}

\preprint{APS/123-QED}

\title{\textbf{Compaction in a deformable porous cylinder with elastic boundaries} 
}%

\author{Richard Mcnair}
\affiliation{Department of Mechanical, Materials and Manufacturing Engineering, University of Nottingham, Nottingham NG7~2RD, UK}
\thanks{Corresponding author}
\email{richard.mcnair@nottingham.ac.uk}

\author{Kerstin Schirrmann}

\author{Anne Juel}
\affiliation{Department of Physics and Astronomy, University of Manchester, Manchester M13~9PL, UK}
\affiliation{Manchester Centre for Nonlinear Dynamics, University of Manchester, Manchester M13~9PL, UK}
\email{anne.juel@manchester.ac.uk}

\author{Igor L. Chernyavsky}
\affiliation{Department of Mathematics, University of Manchester, Manchester M13~9PL, UK}
\affiliation{Maternal and Fetal Health Research Centre, University of Manchester, Manchester M13~9WL, UK}
 \email{igor.chernyavsky@manchester.ac.uk}

\date{\today}

\begin{abstract}
Perfusion of soft materials such as biological tissue or hydrogels is essential for the functioning of organ and laboratory systems such as chromatographic columns and bioreactors. Inspired by these applications, we model fluid-driven compaction in a long, thin cylindrical porous medium bounded by an impermeable \hl{wall, modelled as a thin elastic membrane,} and study how flow regimes relate to elastic parameters. Using a Lagrangian formulation of Darcy flow coupled to small-strain linear elasticity with porosity-dependent permeability and elastic moduli, we perform an asymptotic reduction in the small aspect-ratio limit and obtain a leading-order nonlinear diffusion equation for the porosity, which we solve numerically. Whereas rigid boundaries
produce a compaction plateau, compliant \hl{membranes} exhibit, at most, an intermediate plateau beyond which the flow increases once the imposed pressure becomes comparable to the product of membrane stiffness and initial porosity. When the membrane is less stiff than the porous medium, flow rate can exceed that expected for a rigid medium. A parameter-space map \hl{identifies regimes where the flow rate plateaus before increasing again with applied pressure, distinguishes flow rates} below (sub-Darcy) or above (super-Darcy) the undeformable-medium prediction, and delineates the small-strain domain in which the theory applies. An asymptotic solution for negligible gravity captures the departure from the plateau and yields compact expressions for effective permeability and flow rate.
\end{abstract} 

\maketitle



\section{Introduction}

\begin{figure}
	\centering
		\includegraphics[width=\textwidth]{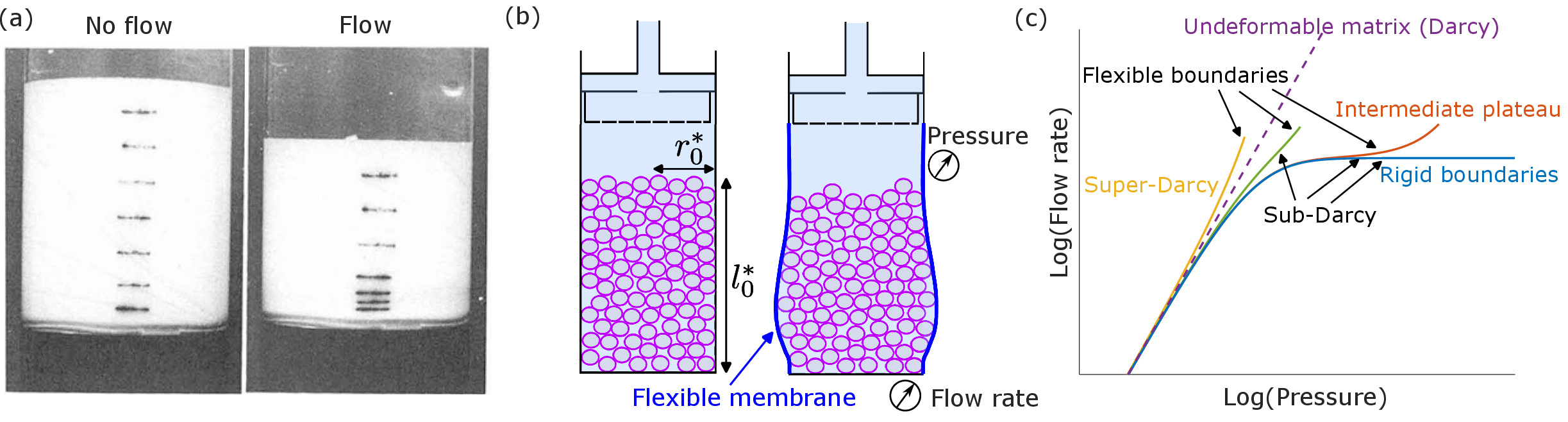}
	\caption{Overview of the problem and qualitative flow-rate regimes. (a) Experimental images adapted from Parker et al.~\cite{parker1987steady} (reproduced by permission of ASME\,$^\text{\copyright}$\,1987): an undeformed sponge marked with equally spaced lines (left) and the sponge during  flow driven by a vertical pressure gradient (right). Line clustering near the outlet reveals nonuniform compaction in a boundary layer. (b) Schematic of the present model, with impermeable but deformable radial boundaries that can displace under flow-induced stresses. (c) Qualitative flow-rate regimes. The purple dashed line shows the undeformable-medium prediction, used as a Darcy benchmark; the blue plateau is the regime identified by Hewitt et al.~\cite{hewitt2016flow}. Flexible radial boundaries can produce  super-Darcy or sub-Darcy responses relative to this benchmark. Within the sub-Darcy class, some cases exhibit an intermediate plateau while others do not.
}
	\label{fig:Introfig}
\end{figure}

Soft, porous materials, such as biological tissues and packed beds of hydrogel beads \cite{bemer2001poromechanics,malandrino2019poroelasticity}, compact under forced perfusion due to fluid pressure gradients that shrink pore space and feed back to the flow (fig.~\ref{fig:Introfig}). This coupling reduces the permeability of a porous medium as its matrix compresses, and beyond a certain point, additional driving pressure yields no extra throughput due to a compaction flow-rate plateau. Generic models of compaction in porous media confined by rigid boundaries were analysed theoretically and experimentally by Parker, \emph{et al.} \cite{parker1987steady}, and more recently by Hewitt, \emph{et al.} \cite{hewitt2016flow} (fig.~\ref{fig:Introfig}a,b). Comprehensive treatments of poroelastic modelling can be found in \cite{cheng2016poroelasticity,coussy2004poromechanics,detournay1993fundamentals}. \hl{The deformation of a poroelastic medium undergoing pressure-driven perfusion can be highly nonuniform. For example, Parker \emph{et al.} }\cite{parker1987steady} \hl{demonstrated experimentally that downward water flow through a cylindrical sponge concentrates compaction in a boundary layer near the outlet, as illustrated in figure}~\ref{fig:Introfig}(a). In many applications the confining boundaries perpendicular to the flow (e.g. organ membranes or polymer housings) are deformable, thereby enabling a hitherto unexplored coupling between the wall mechanics and compaction process. Understanding how the compaction-limited flow is augmented by the boundary compliance is important for interpreting transport in soft media but remains an open question. In this work, we model forced perfusion through a long, thin cylindrical porous medium bounded radially by an unattached impermeable elastic membrane and investigate how the relative wall compliance reorganizes the flow–pressure relationship (fig.~\ref{fig:Introfig}c).

Perfusion in vascularized soft tissues can be idealized as fluid-driven flow through a compliant porous matrix enclosed, at least locally, by a deformable boundary rather than a rigid wall. Adequate perfusion is essential for oxygen and nutrient delivery, yet the same pressure gradients that drive flow can compact the tissue, reduce permeability and impact transport. For example, in the human placenta, maternal blood percolates through a highly deformable villous network \cite{jensen2019blood,abbas2019tissue}. Similar flow–compression trade-offs arise in solid tumors \cite{jin2021poroelasticity}, where elevated interstitial pressure and matrix remodeling can restrict perfusion \cite{stylianopoulos2012causes}, and in thick engineered tissue constructs that rely on direct perfusion before they are fully vascularized \cite{bertassoni2014hydrogel}. In all of these settings the perfused medium and its lateral boundary are mechanically compliant, so understanding when compaction can produce a plateau in flow and when boundary compliance can relieve that plateau is important for understanding perfusion in living tissue and how it can be rescued. 

The role of deformable boundaries is also of interest for the design of packed-bed chromatography columns and perfusion bioreactors \cite{dvorak2024soft,johnson2020packed,keener2002mechanical,podichetty2014modeling}, where the operating geometry (often long, thin cylinders housing deformable porous media) mirrors the configuration analyzed here. \hl{Although conventional chromatography columns are generally designed with rigid housings, more compliant materials are sometimes used} \cite{carr2015lc}\hl{, and radial expansion of column tubes under high operating pressures has been reported} \cite{gritti2018understanding}\hl{. In perfusion bioreactors, compliant polymeric components can also be used to confine porous scaffolds} \cite{piola2013design}. By isolating how radial compliance interacts with pressure-driven flow and compaction in this canonical geometry, our results \hl{provide a framework for understanding the possible effects of boundary compliance in these systems and may inform future designs in which wall stiffness is varied deliberately.} Similar qualitative implications are possible for other porous flows such as confined aquifers \cite{guzy2020state}. Related hydro--mechanical studies reinforce the importance of boundary mechanics. For example, gas migration through bentonite under semi-rigid (transversely compliant) confinement exhibits different pathway formation and breakthrough behaviour than under rigid confinement \cite{guo2024study,liu2025gas}.

Poroelastic regimes with strong flow–structure coupling are often treated with fully nonlinear elastic models, for which numerical solutions are typically the only practical route. MacMinn et al.~\cite{macminn2016large} suggested a hybrid approach that retains nonlinear kinematic links between strain, porosity, and permeability but models the solid’s constitutive law as small-strain linearly elastic, allowing greater analytical traction on the problem. This hybrid approach was adopted with success by Hewitt et al.~\cite{hewitt2016flow}, who coupled a deformation-dependent Kozeny--Carman law to an effective matrix stiffness scaling inversely with liquid volume fraction, and experimentally validated the model using water flow through hydrogel-bead beds, finding the aforementioned flow-rate plateau. Fiori et al.~\cite{fiori2023flow} used models similar to Hewitt’s to understand the effect of periodic loading, and later extended the framework to include solute transport \cite{fiori2025solute}. Variations of Hewitt’s problem were explored by Bouckley et al.~\cite{bouckley2025interplay}, including flow against gravity and external mechanical forcing. Most importantly for the present work, \cite{bouckley2025interplay} generalized effective-stiffness laws proportional to volume fraction raised to a negative exponent, which identify how different exponents select different qualitative flow-rate regimes. Related one-dimensional compaction models for soft, saturated media \cite{paterson2019flow,paterson2022one} couple porosity-dependent permeability to viscoelastic or viscoplastic solid response under fast time-dependent loading, further underscoring the importance of nonlinear kinematics. 

In this study, we follow the same hybrid linear/nonlinear philosophy, but extend it to a two-dimensional, radially compliant cylindrical geometry. We investigate how different material parameters, characterized primarily by the ratio of wall stiffness to \hl{porous-medium stiffness}, correspond to different flow regimes (figure \ref{fig:Introfig}(b,c)). \hl{In this study, we focus primarily on small initial porosities, which allow the nonlinear coupling between deformation and flow to be explored while limiting the magnitude of elastic strains. We recognise that the porosities of many applications motivating this work are considerably larger, and therefore regard the resulting parameter-space maps principally as a means of identifying qualitative flow mechanisms rather than providing quantitative predictions for those systems.} For orientation, figure \ref{fig:Introfig}(c) introduces the qualitative nomenclature used throughout the paper. We take the Darcy flow rate through an undeformable medium as a reference and describe responses below or above this benchmark as `sub-Darcy' or `super-Darcy', respectively. The other curves are schematic and serve to distinguish the classes of flow–pressure response discussed later. When the membrane is much stiffer than the porous medium and the pressure driving the flow, the model we develop recovers the one-dimensional plateauing model of Hewitt et al. \cite{hewitt2016flow}, but diverges from the plateau when the driving pressure rises above the product of membrane stiffness and initial porosity. When the stiffness ratio of membrane to porous medium is only moderately large, we observe a diverging flow rate with no intermediate plateau. When the ratio is small, we observe a flow rate larger than would be expected for a rigid medium.

The paper is organized as follows. In Section \ref{sec:ModandMeth} we present the model, including our choices for the porosity-dependent permeability $k^*(\phi)$ and effective bulk and shear moduli of the porous medium $K^*(\phi)$ and $G^*(\phi)$, which we choose to be the most widely applicable. In Section \ref{sec:res} we present results which show that flexible boundaries can induce a flow rate divergence, contrasting with the plateau behaviour observed with rigid boundaries \cite{hewitt2016flow}, and this is corroborated by a proof-of-principle experiment that predicts that wall compliance suppresses the compaction plateau. We also map out the parameter space in terms of the qualitative behaviour of flow rate with increasing applied pressure. In the discussion section \ref{sec:Disc}, we discuss (especially in light of Bouckley et al. \cite{bouckley2025interplay}) how alternative choices for the permeability function and effective moduli can affect the findings.

\section{Model and methods}
\label{sec:ModandMeth}

We consider steady-state axisymmetric flow and deformation in a tall, thin cylindrical porous medium as shown in Figure \ref{fig:Introfig}(b), which we describe by cylindrical coordinates $(r^*,\theta,z^*)$ with unit vectors $(\mathbf{e}_r,\mathbf{e}_{\theta},\mathbf{e}_z)$, where stars refer to dimensional quantities. In its undeformed configuration, the medium occupies the space $0\leq z^* \leq l_0^*$, $0\leq \theta <2\pi$ and $0\leq r^* \leq r_0^*$. The porosity $\Phi$ of the undeformed reference configuration is uniform and $\phi$ denotes the porosity of the deformed medium. We model the porous medium as saturated by an incompressible liquid, which has dynamic viscosity $\mu^*$ and density $\rho^*$. The solid skeleton is considered to \hl{comprise} an incompressible linear elastic material (infinite bulk modulus) with Young modulus $E_s^*$, and density equal to that of the fluid which is a modeling assumption appropriate for biological tissues and hydrogels, which have a high water content. The medium is bounded radially by an impermeable membrane with Young modulus $E_b^*$, Poisson ratio $\nu_b$, and with wall thickness $t_w^*\ll r_0^*$.

The model presented below is formulated to apply as generally as possible to a wide class of deformable porous media, without assuming specific microstructural details. In doing so, we remain agnostic about the precise mechanical nature of the medium, which may range from granular packings (e.g., hydrogel beads) to continuous elastic matrices (e.g., tissue, sponges). However, we precisely define the regions of parameter space where the model is applicable to granular or connected media, the main difference being that disconnected granular matrices cannot sustain tensile forces. We model the membrane as unattached to the porous medium: it deforms outwards under normal pressure, while tangential slip of the porous medium is allowed. Physically this interaction only makes sense if the porous medium is in contact with the unattached membrane, and so we outline a contact condition which must be met for the model to be valid.

\subsection{Model}

In this subsection, we derive the governing model. We begin by presenting the core conservation laws and the linear elastic constitutive relations. We then nondimensionalize the equations using the small aspect ratio of the tall, thin cylinder. Next, we seek asymptotic expansions for the dependent variables and reduce the system to a single, uncoupled nonlinear diffusion equation governing the leading-order steady-state porosity. Solving this equation determines all other leading-order fields. This study involves a large number of variables and parameters, and so to help the reader, these are summarized in tables in Appendix \ref{app:tables}. Table \ref{tab:dimvarsparams} summarizes dimensional quantities, while the dimensionless leading order variables and model parameters are summarized in Table \ref{tab:keyparamsleadorvars}.

\hl{We adopt a hybrid small-strain poroelastic formulation, following MacMinn et al.} \cite{macminn2016large} \hl{and Hewitt et al.,} \cite{hewitt2016flow} \hl{in which the solid skeleton obeys infinitesimal elasticity while the deformation-dependent kinematics of fluid transport and nonlinear constitutive dependence on porosity are retained. This distinction is important because small strains can produce substantial relative changes in porosity, and hence permeability. We work in Lagrangian coordinates allowing the deformation of the physical domain to be represented through the displacement field while imposing boundary conditions on the fixed reference configuration.}

\subsubsection{Governing equations and boundary conditions}

\hl{At steady state,} the liquid flow is governed by conservation of mass and momentum via Darcy’s law for a divergence-free Darcy velocity. \hl{Throughout, we use the term ``Darcy velocity” to denote the volumetric fluid flux per unit cross-sectional area of the porous medium, which is distinct from the interstitial fluid velocity.} \hl{The Darcy velocity is defined by} the Lagrangian field $\mathbf{U}^*$ and the pressure field is $p^*$. In this setting, Darcy’s law involves the pullback tensor $\mathbf{F}^T \mathbf{F} / J$, where $\mathbf{F}$ is the deformation gradient and $J=\det \mathbf{F}$; a derivation is given in Appendix~\ref{app:pullback}. The medium is assumed to have isotropic permeability $k^*(\phi)$ and to experience \hl{a body force per unit mass} $g^*$ due to gravity such that $\mathbf{g}^* = {\colorbox{yellow}-} g^*\mathbf{e}_z$. 
\begin{align}
	\bm{\nabla}^*\cdot \mathbf{U}^*&=0, \label{eq:consmass1}
\\
	\frac{1}{J}\mathbf{F}^T\mathbf{F}\mathbf{U}^* &= - \frac{k^*(\phi)}{\mu^*}(\bm{\nabla}^*p^*-\rho^* \mathbf{g}^*).\label{eq:darcyslawfulleqn}
\end{align}
The Lagrangian velocity $\mathbf{U}^*$ relates to the Eulerian velocity field $\mathbf{u}^*$ by the Piola transform
\begin{equation}\label{eq:Piola}
    \mathbf{u}^* = \frac{1}{J}\mathbf{F}\mathbf{U}^*,
\end{equation}
where the deformation gradient tensor is given in terms of the solid displacement field $\mathbf{D}^*=[D_r^*,D_z^*]$ by
\begin{equation}
    \mathbf{F} = \mathsf{I} + \bm{\nabla}^*\mathbf{D}^*,
\end{equation}
where $\mathsf{I}$ is the identity tensor. The flow rate through the medium is most easily calculated by integrating the Lagrangian velocity over the undeformed cross-sectional area $A_0^*$
\begin{equation}
    Q^* = {\colorbox{yellow}-} \int_{A_0^*} (\mathbf{U}^*\cdot \mathbf{e}_z) \dee A_0^*.
\end{equation}
The radial component of the Lagrangian Darcy velocity must vanish at the radial boundaries of the initial configuration
\begin{align}
	\mathbf{U}^*\cdot \mathbf{e}_r  &=0, \qquad \text{on } r^*=r_0^* .\label{eq:velboundcondim}
\end{align} 
We drive flow through the cylinder by an imposed pressure $P^*$ above atmospheric pressure \hl{taking the latter to be zero,} such that
\begin{align}
	p^* &= P^*, \qquad \text{on } z^*=l_0^*, \label{eq:pressconddim1}
	\\
	p^*&=0, \qquad \text{on } z^*=0.
\end{align} 

The governing equation for the solid momentum is divergence of the total stress tensor $\bm{\sigma}_{\mathrm{tot}}^*$ equal to the specific gravitational force on the cylinder. The total stress tensor is modelled using Terzaghi's principle as the sum of an effective stress tensor $\bm{\sigma}_{\mathrm{eff}}^*$ plus an isotropic component proportional to fluid pressure. \hl{We adopt a compression-positive convention for stress and the standard tension-positive convention for strain. Consequently, the linear elastic constitutive relation contains an overall negative sign, such that compressive strains produce positive effective stresses, following Hewitt et al.}~\cite{hewitt2016flow}. The effective stress tensor relates to the strain tensor $\bm{\varepsilon}$ according to a standard linear elastic formulation with nonlinearities entering in via deformation dependent bulk and shear moduli $K^*(\phi)$ and $G^*(\phi)$. \hl{These moduli depend on the porosity of the deformed configuration as, following Hewitt et al.} \cite{hewitt2016flow}, \hl{we seek to model deformation-dependent changes in the stiffness of the medium and how these feed back into the flow.} These effective moduli should capture the bulk and shear moduli of the incompressible solid phase as $\phi$ goes to zero ($\infty$ and $E_s^*/3$ respectively for an incompressible solid). For now we will leave the effective moduli in their general form and wait until subsection \ref{sec:modmethConstAssump} to choose a model. \hl{We formulate the solid mechanics using infinitesimal elasticity, so that the stresses appearing in the equilibrium and constitutive equations are linearised stresses, with distinctions between finite-deformation stress measures neglected at the order of approximation considered.} The solid mechanics equations are
\begin{align}
     \bm{\nabla}^*\cdot \bm{\sigma}^*_{\mathrm{tot}} &= \rho^*\mathbf{g}^* ,\label{eq:cauchysfulleqn}
\\
	\bm{\sigma}^*_{\mathrm{tot}} &= \bm{\sigma}^{*}_\mathrm{eff}+p^*\mathsf{I}, \label{eq:Terzaghi}
\\ 
	\bm{\sigma}^*_{\mathrm{eff}} &= - K^*(\phi)\text{Tr}(\bm{\varepsilon})\mathsf{I}-2G^*(\phi)\left(\bm{\varepsilon}-\frac{1}{3}\text{Tr}(\bm{\varepsilon})\mathsf{I}\right) .\label{eq:effstressfulleq}
\end{align}
The porous matrix, deforming due to gravity and pore-pressure gradients, transmits stresses to the bounding membrane and induces its deformation. We model this coupling with the \KD thin-shell equations, which relate the shell’s normal (radial) displacement to the applied loads \cite{timoshenko1959theory} \cite{xue2013extension}. In the axisymmetric setting, this reduces to a single boundary condition. The total \hl{radial traction from} the porous medium balances a hoop-tension restoring term for the radial displacement $D_r^*$, proportional to the membrane’s Young’s modulus $E_b^*$ and thickness $t_w^*\ll r_0^*$. A bending contribution enters through a fourth derivative of the displacement scaled by the flexural rigidity $E_b^*t_w^{*3}/12(1-\nu_b^2)$ where $\nu_b$ is the membrane's Poisson ratio. Along the membrane, the porous medium is allowed to slip freely. Therefore,
\begin{align}
	\frac{E_b^*t_w^{3*} }{12(1-\nu_b^2)} \frac{\dee^4 D^*_r}{\dz^{*4}} + \frac{E_b^*t_w^*}{r_0^{*2}} D_r^* &= \mathbf{e}_{r}\cdot \bm{\sigma}^*_{\mathrm{tot}}\cdot \mathbf{e}_{r} , \qquad\text{on } r^*=r_0^*,  \label{eq:dimensionalFvK}
    \\
	\mathbf{e}_r\cdot\bm{\sigma}^*_{\mathrm{tot}}\cdot\mathbf{e}_z&=0, \qquad \text{on } r^*=r_0^*.
\end{align} 
To fully isolate the role of wall compliance, we impose clamped boundary conditions on \eqref{eq:dimensionalFvK}, effectively imposing infinitely stiff collars at the top and bottom of the porous medium. This yields
\begin{align}
    D_{r}^* &= 0, \qquad \text{on } z^* = 0 ,l_0^*
    \\
    \frac{\partial D_r^*}{\partial z^*} &= 0, \qquad \text{on } z^* = 0 ,l_0^* . \label{eq:boundconderivDr}
\end{align} 
The strain relates to the displacement by linearised relations. \hl{Consistent with the infinitesimal-elasticity formulation, we use the linearised strain tensor, neglecting terms quadratic in the displacement gradients that would arise in a finite-strain formulation.} A small-strain identity links volumetric strain to the Jacobian, $J^{-1}=1-\mathrm{tr}\,\bm{\varepsilon}+O(\varepsilon^{2})$, which we use because it yields the linear expression $\mathrm{tr}\,\bm{\varepsilon}\approx (\phi-\Phi)/(1-\Phi)$. \hl{These conditions constrain the radial displacement and slope of the membrane at both ends, but do not restrict the axial displacement of the porous medium, which is free to shorten under flow-induced compaction.} Finally the definition of $J$ in terms of porosity is an exact geometrical relation for an incompressible solid. Thus,
\begin{align}
\bm{\varepsilon} &= \frac{1}{2}\left(\bm{\nabla}^*\mathbf{D}^{*}+\bm{\nabla}^*\mathbf{D}^{*\mathsf{T}}\right) ,\label{eq:straindisp}
	\\
	\text{Tr}\left(\bm{\varepsilon}\right) &= 1-\frac{1}{J} , \label{eq:traceepsilon2}
    \\
    J &= \frac{1-\Phi}{1-\phi}. \label{eq:mainJacobianeq}
\end{align}
At the top of the cylinder, the total-stress must equal the applied pressure, hence through \eqref{eq:Terzaghi} the effective stress must be zero. As radial clamping and axisymmetry leaves only one free coordinate direction, zero effective stress here is equivalent to zero strain. Hence through \eqref{eq:mainJacobianeq} we have
\begin{equation}
    \phi=\Phi  \qquad \text{on } z^*=l_0^*.\label{eq:totstresscont}
\end{equation} 

\subsubsection{Contact condition and applicability to granular media}

\hl{Although the porous skeleton exerts an outward compressive force on the unattached membrane, the membrane also deforms in response to fluid pressure. For sufficiently compliant membranes, this pressure-driven expansion may exceed the radial expansion of the porous skeleton, potentially compromising the assumption that the medium remains in contact with the membrane. Since the membrane is unattached, we impose a condition of compressive total stress at the boundary} (\cite{hosseinkhan2021biot,banz2025contact} ). \hl{We therefore restrict our solutions to those satisfying the contact condition}
\begin{equation}\label{eq:dimcontactcondition}
    \mathbf{e}_r \cdot \bm{\sigma}^*_{\mathrm{tot}} \cdot \mathbf{e}_r \geq 0 \qquad \text{on } r^* = r_0^*.
\end{equation}
\hl{Equation} \eqref{eq:dimcontactcondition} \hl{thus provides a practical criterion for identifying solutions consistent with the assumed contact between the porous medium and the membrane.}

For granular media, we must further constrain the applicability of the model as they cannot sustain tensile forces anywhere (not just at the wall) due to their unconnected nature. This condition is equivalent to constraining the effective stress to be compressive everywhere. \hl{ In the present configuration, tensile effective stresses may arise in the radial direction, so we require}
\begin{equation}
  \mathbf{e}_r \cdot \bm{\sigma}^*_{\mathrm{eff}} \cdot \mathbf{e}_r \geq 0,  \qquad \text{everywhere for granular media,}
\end{equation}
and we reject applications of the model to granular media for regions of the parameter space where this condition is not met.

\subsubsection{Non-dimensional model}

We nondimensionalize equations~\eqref{eq:consmass1} to~\eqref{eq:totstresscont} using the geometry and material parameters of the system. Vertical lengths and displacements are scaled by the cylinder height \( l_0^* \), while radial lengths and displacements are scaled by \( r_0^* = \epsilon l_0^* \), where \( \epsilon \ll 1 \) is the aspect ratio of the cylinder. All stress components, pressures, and effective elastic moduli are nondimensionalized by the Young modulus of the solid skeleton \( E_s^* \). The permeability is written in the form \( k^*(\phi) = \overline{k}^* k(\phi) \), where \( \overline{k}^* \) is a characteristic reference value and \( k(\phi) \) is the dimensionless permeability function. Radial and vertical flow velocities are scaled by \( \overline{k}^* E_s^*/\mu^* r_0^* \) and \( \overline{k}^* E_s^*/\mu^* l_0^* \), respectively.  A summary of the nondimensionalization scheme is
\begin{align}\label{eq:nondimscheme}
    z &= \frac{z^*}{l_0^*} \qquad r= \frac{r^*}{\epsilon l_0^*} \qquad p = \frac{p^*}{E_s^*} \qquad k(\phi) = \frac{k^*(\phi)}{\bar{k}^*} \qquad K(\phi) = \frac{K^*(\phi)}{E_s^*} \nonumber
    \\
    G(\phi) &= \frac{G^*(\phi)}{E_s^*} \qquad U_z = \frac{ U_z^* \mu^* l_0^*}{\bar{k}^*E_s^*}  \qquad 
    D_r = \frac{D_r^*}{\epsilon l_0^*} \qquad D_z = \frac{D_z^*}{l_0^*} \qquad
    U_r = \frac{\epsilon U_r^* \mu^* l_0^*}{\bar{k}^*E_s^*} .
    \end{align}
    This yields a non-dimensional model in five parameters, the imposed fluid pressure $\mathcal{P}$, the gravitational pressure $\Lcal$, the ratio of membrane to porous medium stiffness $\Scal$ and the bending parameter $\Tcal$, defined as
    \begin{equation}
        \Pcal  = \frac{P^*}{E_s^*}, \qquad 
        \Lcal  = \frac{\rho^* g^* l_0^*}{E_s^*},  \qquad
        \Scal  = \frac{E_b^* t_w^*}{E_s^* r_0^*}, \qquad
        \Tcal  = \frac{\epsilon^4 t_w^{*2}}{12r_0^{*2}(1-\nu_b^2)},
    \end{equation}
    along with the initial porosity $\Phi$.

We seek asymptotic series approximations for the dependent variables $U_r = U_{r0}+\epsilon^2 U_{r1}+\dots$, $U_z = U_{z0}+\epsilon^2 U_{z1}+\dots$, $D_r = D_{r0}+\epsilon^2 D_{r1}+\dots$, $D_z = D_{z0}+\epsilon^2 D_{z1}+\dots$, and $\phi = \phi_{0}+\epsilon^2 \phi_{1}+\dots$. \hl{The asymptotic expansions are written in powers of $\epsilon^2$, since substitution of expansions in integer powers of $\epsilon$ into the governing equations and boundary conditions shows that all $O(\epsilon)$ corrections to the leading order variables vanish.} While the leading order variables $\phi_0$, $D_{r0}$ and $D_{z0}$ are formally considered $O(1)$ in these expansions, their physical magnitudes are all constrained to be small (much less than 1) to satisfy the linear elasticity assumption. The displacements $D_{r0}$ and $D_{z0}$ are small to produce small strains, and $\Phi-\phi_0$ is small to reflect small volume changes. These physical constraints do not affect the asymptotic ordering as higher order-terms are suppressed by factors of $\epsilon^2 \ll1$. If we were to include higher order terms in $\epsilon$, we would have to consider asymptotic relationships between the sizes of strains and $\epsilon$. 

As shown in Appendix \ref{app:simp}, the system of equations and boundary conditions \eqref{eq:consmass1} to \eqref{eq:totstresscont} and scaling relations \eqref{eq:nondimscheme} can be combined to leading order in $\epsilon^2$ to give the coupled equations
\begin{equation}\label{eq:Qz0gen01}
	 \frac{\dee}{\dz}\left[  \frac{k(\phi_0)(1+\chi(z))^4(1-\phi_0)}{1-\Phi}
	 \frac{\partial}{\partial z} \left[ M(\phi_0)\left(\frac{\phi_0-\Phi}{1-\Phi} \right) -4G(\phi_0)\chi(z)\right]  \right] = 0 .
\end{equation}
and
\begin{equation}\label{eq:nonDFvk2}
\mathcal{T}\mathcal{S} \frac{d^4 \chi(z)}{dz^4} + \mathcal{S} \chi(z) = \mathcal{P} + \mathcal{L}(1 - z) + 2G(\phi_0)\left( \frac{\phi_0 - \Phi}{1 - \Phi} - 3\chi(z) \right).
\end{equation}
where $\chi(z) = D_{r0}/r$ is the nondimensional radial profile, and $M(\phi_0)=K(\phi_0)+4G(\phi_0)/3$ is the nondimensional effective longitudinal modulus of the solid phase (often called the p-wave modulus). These are to be solved according to boundary conditions of 
\begin{equation}\label{eq:FVKBoundConds}
    \chi(z) = \frac{\dee \chi(z)}{\dz} = 0 \qquad \text{on } z=0,1
\end{equation}
and 
\begin{equation}\label{eq:phisolboundconds}
    \phi_0(1) = \Phi \qquad \phi_0(0) = \varphi,
\end{equation}
where $\varphi$ is found through the solution of
\begin{equation}\label{eq:varphieqGen}
\mathcal{P} + \mathcal{L} + M(\varphi) \left( \frac{\varphi - \Phi}{1 - \Phi} \right) = 0.
\end{equation}
The Lagrangian Darcy velocity through the cylinder is then given by
\begin{equation}\label{eq:VelTimeDep}
 U_{z0} = - \frac{k(\phi_0)(1+\chi(z))^4(1-\phi_0)}{1-\Phi}\frac{\partial}{\partial z} \left[ M(\phi_0)\left(\frac{\phi_0-\Phi}{1-\Phi} \right) -4G(\phi_0)\chi(z)\right]  .
\end{equation} 
and the total flow rate  is given by
\begin{equation}\label{eq:flowrateequation}
    Q_{0} = \colorbox{yellow}{-}\pi U_{z0}.
\end{equation}
 The scaling for the nondimensional volumetric flow rate \eqref{eq:flowrateequation} is $ \epsilon^2 \bar{k}^* E_s^* l_0^*/\mu^*$. The contact condition \eqref{eq:dimcontactcondition} becomes
\begin{equation}\label{eq:contcondnond}
    \Pcal+\Lcal(1-z) +2G(\phi_0) \left( \frac{\phi_0-\Phi}{1-\Phi} \right) -6G(\phi_0) \chi(z) \geq 0.
\end{equation}

 \subsubsection{Simplification of the coupled equations and boundary layer solution for the membrane equation}

 A simplification of the equations \eqref{eq:nonDFvk2} and \eqref{eq:Qz0gen01} can be achieved by noticing that the terms dependent on $\phi_0$ in \eqref{eq:nonDFvk2} are always $O(\Phi)$ smaller than the forcing $\mathcal{P}+\Lcal(1-z)$. Writing $G(\phi_0)\approx G(\Phi)+G'(\Phi)(\phi_0-\Phi)+\dots$ and removing terms dependent on $\phi_0$ leaves
 \begin{equation}\label{eq:nonDFvk3}
	\mathcal{T}\mathcal{S} \frac{\dee^4 \chi(z)}{\dz^{4}}  + (\mathcal{S}+6G(\Phi))\chi(z) = \mathcal{P}+\mathcal{L}(1-z) ,
\end{equation} 
which can be solved directly and the solution for $\chi(z)$ substituted into \eqref{eq:Qz0gen01} uncoupling the system and reducing unnecessary computational cost. This simplification is rigorously justified in Appendix \ref{app:assump} as well as a comparison between this simplification and full solutions of \eqref{eq:nonDFvk2} and \eqref{eq:Qz0gen01}, but for the rest of the study we focus on the simplified system \eqref{eq:nonDFvk3} and \eqref{eq:Qz0gen01}.

Equation \eqref{eq:nonDFvk3} has a full analytical solution according to boundary conditions \eqref{eq:FVKBoundConds}, but this solution is difficult to work with numerically because \eqref{eq:nonDFvk3} is a singular differential equation because $\mathcal{T}\ll1 $. We therefore solve \eqref{eq:nonDFvk3} according to \eqref{eq:FVKBoundConds} via matched asymptotics as shown in Appendix \ref{app:radialprofilesol} giving solution
\begin{equation}\label{eq:F(z)}
  \chi(z) = \frac{
\mathcal{P}+\mathcal{L}(1-z) + \chi_b(z)}{  \mathcal{S}+6G(\Phi)} 
\end{equation}
where
\begin{multline}\label{eq:FbandOmega}
    \chi_b(z) = - e^{-\Omega z}(\mathcal{P}+\mathcal{L})\left(\cos{(\Omega z)} +\left(1- \frac{\mathcal{L}}{\Omega(\mathcal{P}+\mathcal{L})}\right) \sin{(\Omega z)} \right)
    \\
-  e^{-\Omega (1-z)}\mathcal{P}\left(\cos{(\Omega(1-z))} +(1 \colorbox{yellow}{+} \frac{\mathcal{L}}{\Omega \mathcal{P}}) \sin{(\Omega (1-z))} \right), \qquad    \Omega = \left(\frac{(\mathcal{S}+6G(\Phi))}{4\mathcal{S}\mathcal{T}} \right)^{1/4} \gg1
\end{multline}

\subsection{Constitutive models}
\label{sec:modmethConstAssump}

The previous subsection yields an uncoupled differential equation for $\phi_0$ for a general choice of effective moduli $M(\phi_0)$ and $G(\phi_0)$ and permeability $k(\phi_0)$. To make  progress we now need to make constitutive assumptions about the effective elastic moduli and the permeability. We will choose the most widely representative models. These will be the Mackenzie model of effective bulk and shear moduli given in full in Appendix \ref{app:MackenzieModel} and derived in \cite{mackenzie1950elastic}, which are found by upscaling the equations of linear elasticity around spherical pores; it is considered a widely applicable model for media with small porosity \cite{cheng2016poroelasticity}, and is a direct two-dimensional extension of the effective modulus used in \cite{hewitt2016flow}. When we assume incompressibility of the solid phase, Mackenzie's relations become (to $O(1)$)
\begin{equation}\label{eq:BulkandShearModused}
	M(\phi) = \frac{4}{9\phi}, \qquad G(\phi) = \frac{1}{3} .
\end{equation}
The permeability function we choose is the Kozeny-Carman model
\begin{equation}
	k(\phi) = \frac{\phi^3}{(1-\phi)^2}.
\end{equation} \hl{We emphasise that these constitutive laws are particular choices within the more general formulation of equation} \eqref{eq:Qz0gen01}\hl{, which permits alternative porosity-dependent permeability and elastic moduli. The influence of different constitutive laws on the resulting flow regimes is discussed in Section} \ref{sec:Disc}. In the case of the Mackenzie model, \eqref{eq:varphieqGen} reduces to 
\begin{equation}\label{eq:varphieqn}
 \varphi = \frac{\Phi}{1+9(\mathcal{P}+\mathcal{L})(1-\Phi)/4}.
\end{equation}
The nonlinear diffusion equation \eqref{eq:Qz0gen01}, under the Mackenzie and Kozeny-Carman models becomes
\begin{equation}\label{eq:NonlinDiff2}
		\frac{\dee}{\dz}\left[  \frac{\Phi\phi_0(1+\chi)^4}{(1-\Phi)(1-\phi_0)} \frac{\dee \phi_0}{\dz} - \frac{3\phi_0^3(1+\chi)^4}{(1-\phi_0)}\frac{\dee \chi}{\dz} \right] = 0 ,
\end{equation}
with spatially constant flow rate 
\begin{equation}\label{eq:flowrate}
Q_{0} =    \frac{4\pi\Phi\phi_0(1+\chi)^4}{9(1-\Phi)^2(1-\phi_0)} \frac{\dee \phi_0}{\dz} -\frac{4\pi\phi_0^3(1+\chi)^4}{3(1-\phi_0)(1-\Phi)}\frac{\dee \chi}{\dz}.
\end{equation}

From \eqref{eq:straindisp}, the strains of the porous medium in the model are
\begin{equation}\label{eq:solutionstrains}
  \bm{\varepsilon}  = \begin{pmatrix}
     \chi(z) & 0 & \epsilon \frac{r}{2} \frac{\dee \chi(z)}{\dz}
     \\
     0& \chi(z)
      & 0\\ \epsilon \frac{r}{2} \frac{\dee \chi(z)}{\dz} & 0 & \frac{\phi_0-\Phi}{1-\Phi} - 2\chi(z) 
  \end{pmatrix}.
\end{equation}
 The component of strain with the largest magnitude is $\varepsilon_{zz}$. To remain consistent with the small strain approximation we will reject any solution where $\max{\left|\varepsilon_{zz}\right|}>0.20$. Solutions are obtained by solving \eqref{eq:NonlinDiff2} according to $\phi(1)=\Phi$ and $\phi(0)=\varphi$ using a Newton scheme, the details of which are in Appendix \ref{app:Numerics}.

\subsection{Asymptotic solution in negligible gravity limit}
\label{sec:Asymp}

The numerical solution of \eqref{eq:NonlinDiff2} determines the full steady response, but does not immediately reveal how wall compliance modifies the compacted, rigid-wall state. To isolate this mechanism, we consider pressure-driven flow in the negligible-gravity limit, $\mathcal{L}=0$, and derive an asymptotic approximation in a regime where the wall deformation remains small but has an appreciable effect on the porosity field. Defining
\begin{equation}
    \gamma =2 \mathcal{P}/(\mathcal{S}+2),
\end{equation}
we consider the asymptotic regime
 \begin{equation}\label{eq:gammacond}
   1 \gg \gamma \gg \Phi,
 \end{equation}
together with $\Omega \gg1$, yielding
\begin{equation}
	\frac{ \dee \chi(z)}{\dz} = \gamma \Omega \left( e^{-\Omega z} \sin{(\Omega z)} -e^{-\Omega(1-z)}\sin{(\Omega(1-z))} \right) .
\end{equation}
The resulting approximation provides explicit expressions for the porosity, effective permeability, and flow rate, and identifies how wall compliance produces departure from the rigid-wall compaction plateau. We first rescale the porosity by its reference value, writing $\tilde{\phi} = \phi_0/\Phi$, so that \eqref{eq:NonlinDiff2} becomes 
\begin{equation}\label{eq:LeadOrdEq}
	\frac{\dee}{\dz}\left[  \frac{\tilde{\phi}(1+\chi(z))^4}{(1-\Phi)(1-\Phi\tilde{\phi})} \frac{\dee \tilde{\phi}}{\dz} - \frac{3\tilde{\phi}^3(1+\chi(z))^4}{(1-\Phi\tilde{\phi})}\frac{\dee \chi}{\dz} \right] = 0 .
\end{equation}
We look for an asymptotic series $\tilde{\phi} = \tilde{\phi}_0 + \gamma \tilde{\phi}_1+\dots$, then \eqref{eq:LeadOrdEq} becomes 
\begin{equation}\label{eq:leadordeq1}
	\frac{\dee}{\dz}\left( \tilde{\phi}_0 \frac{\dee \tilde{\phi}_0}{\dz} \right) = 0  ,
\end{equation}
\begin{equation}\label{eq:leadordeq2}
	\frac{\dee}{\dz} \left( \frac{\dee}{\dz }\left(\tilde{\phi}_0\tilde{\phi}_1\right) - 3 \tilde{\phi}^3_0 \Omega\left( e^{-\Omega z} \sin{(\Omega z)} -e^{-\Omega(1-z)}\sin{(\Omega(1-z))} \right) \right) = 0 ,
\end{equation}
where we are missing terms in \eqref{eq:leadordeq2} which will integrate to $O(1/\Omega)$, and with boundary conditions
\begin{equation}
	\tilde{\phi}_0 = \varphi/\Phi , 1 \qquad \text{on } z = 0 ,1,
\end{equation}
\begin{equation}
	\tilde{\phi}_1 = 0 \qquad \text{on } z = 0,1,
\end{equation}
where $B=\varphi/\Phi$. As shown in Appendix \ref{app:asympsol}, we solve these equations and relate the harmonic mean permeability to flow rate to find the solution for the porosity
\begin{multline}\label{eq:asympporo}
	\tilde{\phi} = \sqrt{(1-B^2)z+B^2} + 
	\frac{3 \gamma}{2} \Bigg( \frac{B^3+(1-B^3)z}{\sqrt{(1-B^2)z+B^2}}
    \\
    -\,e^{-\Omega z} ((1-B^2)z+B^2)[\sin(\Omega z)+\cos(\Omega z)]
	-\,e^{-\Omega(1-z)} ((1-B^2)z+B^2)[\sin(\Omega(1-z))+\cos(\Omega(1-z))] \Bigg).
\end{multline}
The equation for Lagrangian flow rate, given that we already know the vertical Darcy velocity is a spatial constant, is
\begin{equation}
	\frac{(1-\Phi)Q_0}{(1-\phi_0)(1+\chi(z))^4} = k(\phi_0) \pi\frac{\dee p}{\dz}.
\end{equation}
We can divide by $k(\phi_0)$ and integrate over the domain, giving the flow rate proportional to the harmonic mean permeability. The resulting flow rate is
\begin{equation}\label{eq:asympflowrate}
	Q_{asymp} = \frac{\Phi^3\pi B(1+B)\mathcal{P}}{2(1-\Phi)} \left(  1 + 3\gamma \left( 2+ \frac{B^2}{1+B} \right) - \frac{\Phi}{1-\Phi} \left( \frac{2}{3} \left(1+ \frac{B^2}{1+B} \right)-1+(1-\Phi) \frac{B\ln{B^2}}{2(1-B)} \right)  \right).
\end{equation}

\section{Results}
\label{sec:res}

\begin{figure}
	\centering
\begin{subfigure}{0.48\textwidth}
\includegraphics[width=\textwidth]{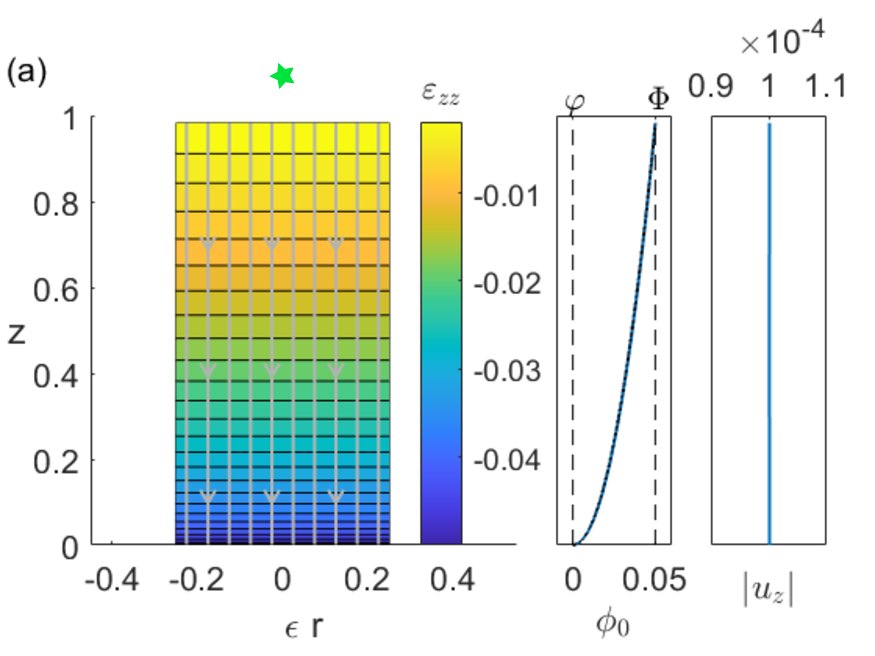}
\end{subfigure}
\begin{subfigure}{0.48\textwidth}
\includegraphics[width=\textwidth]{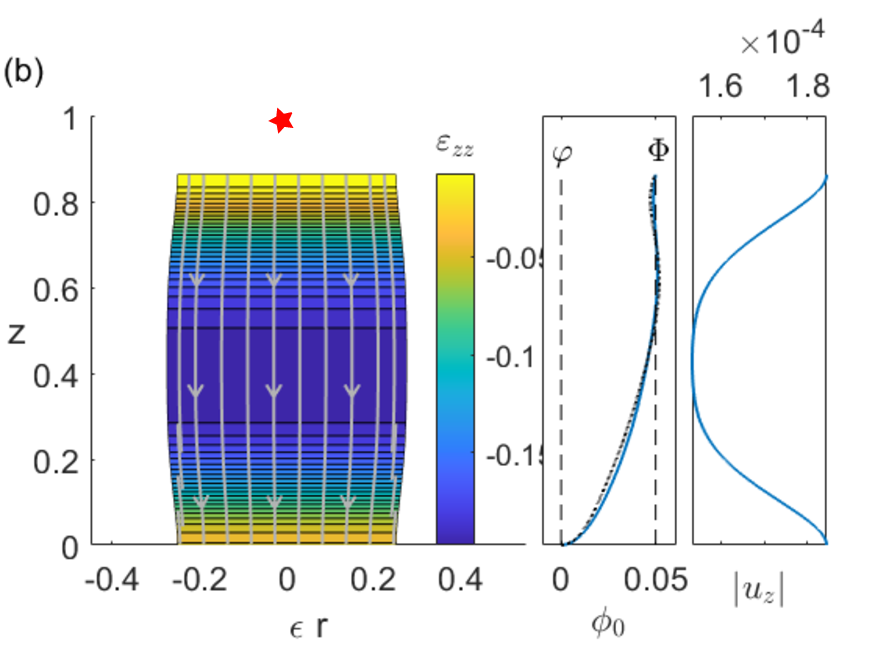}
\end{subfigure}
\begin{subfigure}{0.48\textwidth}
\includegraphics[width=\textwidth]{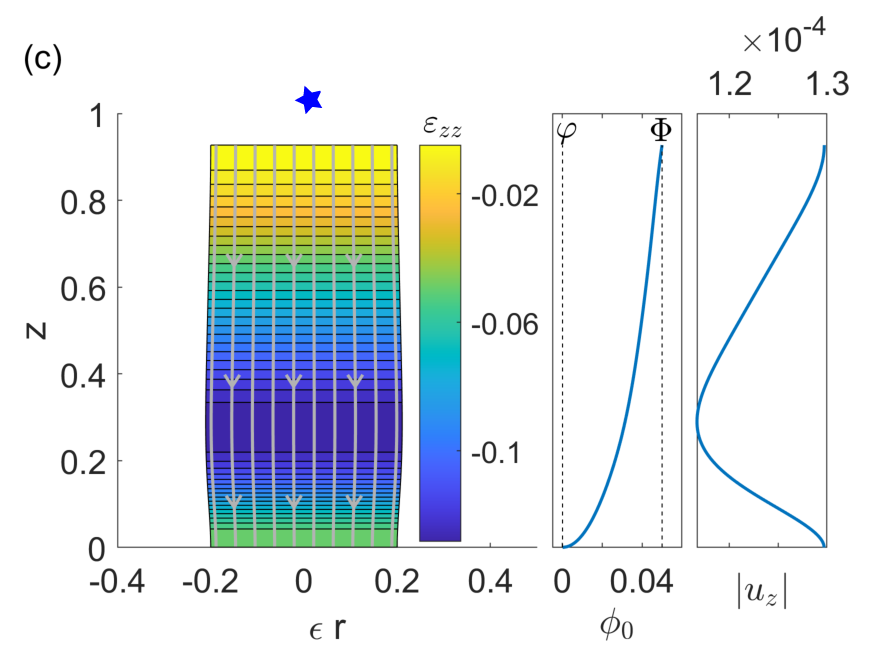}
\end{subfigure}
\begin{subfigure}{0.48\textwidth}
\includegraphics[width=\textwidth]{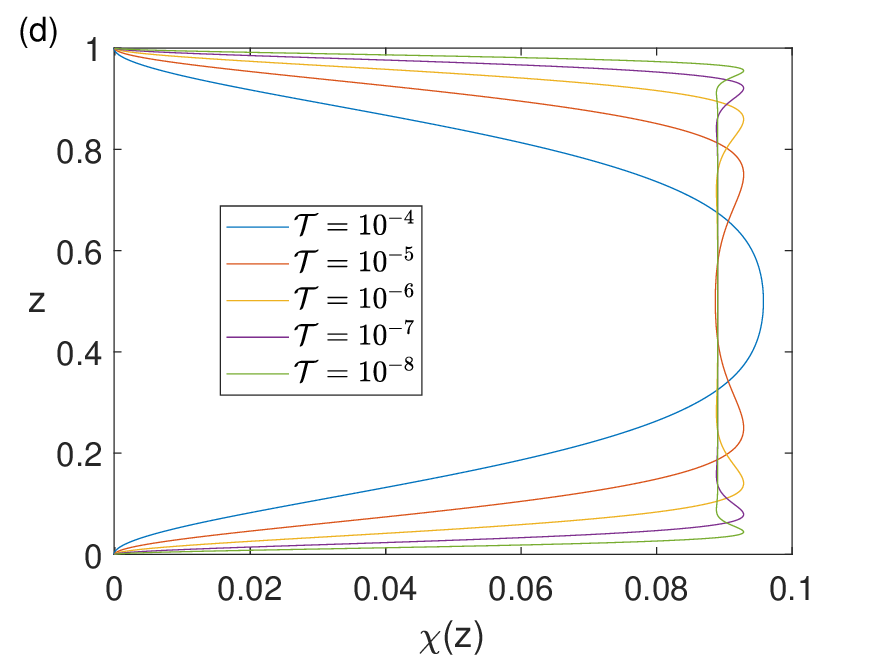}
\end{subfigure}
\caption{Examples of deformation, strain, porosity, and Eulerian Darcy velocity in a deformed cylinder for $\Phi=0.05$ \hl{and $\mathcal{T}=10^{-4}$}. (a--c) Deformed cylinders coloured by axial strain $\varepsilon_{zz}$; grey curves are flow streamlines. Adjacent profiles show the porosity $\phi_0$ and the magnitude of the axial Eulerian Darcy velocity $|u_z|$ plotted against the same $z$-coordinate; dotted curves in (a,b) show the asymptotic porosity solution \eqref{eq:asympporo}. The parameter values are $(\Pcal,\Scal,\Lcal)=(890,10^7,0)$ in (a), $(890,10^4,0)$ in (b), and \hl{$(0,10^4,800)$} in (c). Coloured stars mark the corresponding locations in the parameter maps of Fig.~\ref{fig:Regimes}. (d) Radial displacement $\chi(z)$ for the solution with $(\Pcal,\Scal,\Lcal)=(890,10^4,0)$ and several values of $\Tcal$, as indicated in the legend. 
}
	\label{fig:straindef}
\end{figure}

\subsection{Wall compliance redistributes compaction}

We begin by examining how axial compaction, radial expansion, and flow organize themselves in steady state for representative parameter sets, as shown in Fig.~\ref{fig:straindef}. Panels~(a--c) plot the deformed cylinder coloured by axial strain $\varepsilon_{zz}$, with streamlines of the Eulerian Darcy velocity overlaid in gray; to the right of each panel we show the corresponding axial profiles of porosity and vertical Darcy velocity. Figure \ref{fig:straindef}(a) corresponds to a pressure-driven case with $\Pcal = 8.9 \times 10^{2}$, negligible gravity $(\Lcal = 0)$, and a very stiff membrane $(\Scal = 10^{7})$, so that the radial boundary is effectively rigid. The deformation is then predominantly axial: $\varepsilon_{zz}$ is most compressive near the outlet and the porosity drops sharply in that outlet-adjacent region consistent with the classic compaction boundary layer reported for rigid systems.

Figure~\ref{fig:straindef}(b) shows the same pressure load $(\Pcal = 8.9 \times 10^{2},\, \Lcal = 0)$ but with a more compliant boundary $(\Scal = 10^{4})$, so that the membrane expands outwards radially. In this regime the cylinder expands radially in the middle of its length, reducing the axial compression locally. As a result the porosity remains close to its inlet value over a much larger fraction of the column, and the vertical Darcy velocity is reduced in the bulging region where the cross-section is largest. The dotted black curve in the porosity plot is the asymptotic prediction \eqref{eq:asympporo}, which closely tracks the numerical solution.

\hl{When $\mathcal{P}=0$, gravity alone can drive both deformation and flow through the saturated porous medium, provided fluid is supplied at the inlet. The resulting steady-state deformation therefore consists of gravitational `sagging' of the medium, without an imposed driving pressure difference between inlet and outlet.} Figure~\ref{fig:straindef}(c) isolates gravity-driven flow with \hl{$\Lcal = 800$}, $\Pcal = 0$, and $\Scal = 10^{4}$. Here the strain and bulging become asymmetric. The lower part of the column carries most of the compaction, and the membrane bows outward more strongly toward the bottom where the hydrostatic pressure is highest. This generates a corresponding vertical asymmetry in porosity and Darcy velocity. For panels~(a--c) we fix the dimensionless bending parameter at $\Tcal = 10^{-4}$.

Finally, figure~\ref{fig:straindef}(d) shows how the radial displacement profile $\chi(z)$ depends on the bending parameter $\Tcal$ for a representative pressure-driven case $(\Pcal = 8.9 \times 10^{2},\, \Lcal = 0,\, \Scal = 10^{4})$. Decreasing $\Tcal$ weakens bending resistance and produces a flatter, more plug-like radial expansion in the interior, with the membrane curvature confined to narrow boundary layers near the clamped ends. This confirms that $\Tcal$ primarily controls how sharply the wall returns to its undeformed radius at $z=0$ and $z=1$, while leaving the bulk expansion amplitude largely unchanged.

\subsection{Flow-rate plateau and compliance-induced recovery}

\begin{figure}
	\centering
	\begin{subfigure}{0.495\textwidth}
		\includegraphics[width=\textwidth]{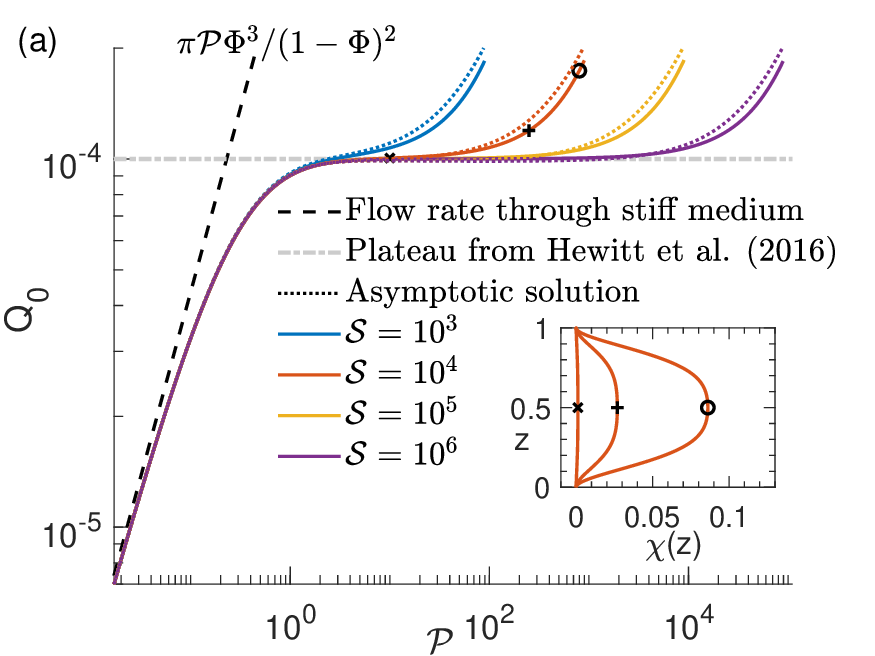}
	\end{subfigure}
	\begin{subfigure}{0.495\textwidth}
		\includegraphics[width=\textwidth]{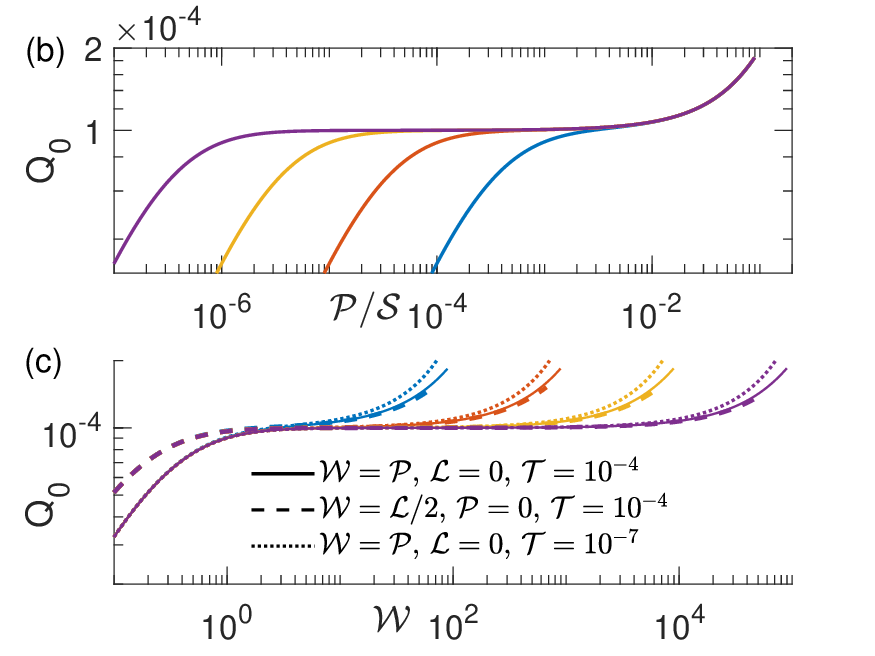}
	\end{subfigure}
	\caption{Logarithmically scaled plots of dimensionless flow rate $Q_{0}$ versus dimensionless driving pressure with undeformed porosity $\Phi=0.05$ for cases where the membrane is much stiffer than the porous skeleton. Curves terminate when $\max{|\varepsilon_{zz}|}>0.2$.
(a) $Q_{0}$ against $\mathcal{P}$ for values of stiffness ratio $\mathcal{S}$ shown in the legend (colour-coded values of $\Scal$ shown in legend apply to (a)-(c)) with $\mathcal{T}=10^{-4}$ and $\mathcal{L}=0$. Solid lines: \eqref{eq:flowrate} evaluated using $\phi_0$ from \eqref{eq:NonlinDiff2}. 
Dashed black: Darcy flow rate in an undeformable medium. Grey dash--dotted: plateau solution from \cite{hewitt2016flow}. 
Dotted: asymptotic approximation \eqref{eq:asympflowrate}. Inset: radial profiles $\chi(z)$ for $\mathcal{S}=10^{4}$ at $\mathcal{P}=10,\,250,$ and $800$, as marked on the main plot. 
(b) Same $\mathcal{S}$ values as in (a), with the horizontal axis scaled by $\mathcal{S}$, showing collapse at large $\mathcal{P}$. 
(c) Curves from (a) (solid) together with solutions for $\mathcal{P}=0$ plotted against $\mathcal{L}/2$ (dashed), demonstrating collapse of the two families. 
Dotted: solutions for the parameters of (a) with $\mathcal{T}=10^{-7}$, indicating only a weak dependence on $\mathcal{T}$.   
}
	\label{fig:Mainresults2}
\end{figure}

We next quantify how the steady volumetric flow rate responds to driving pressure in the regime where the membrane is much stiffer than the porous skeleton, so that the system still exhibits an intermediate flow-rate plateau. Figure~\ref{fig:Mainresults2}(a) shows the non-dimensional flow rate $Q_0$ against the imposed pressure $\Pcal$ for several stiffness ratios $\Scal$ at zero gravity $\Lcal = 0$, bending parameter $\Tcal = 10^{-4}$, and undeformed porosity $\Phi = 0.05$. For small $\Pcal$, all curves follow the Darcy prediction for a rigid medium: the porous matrix is essentially uncompressed, and the flow rate is proportional to  pressure. As $\Pcal$ increases, the porous matrix compacts axially near the outlet and the flow rate ceases to grow, producing a plateau consistent with the one-dimensional model and experiments of \cite{hewitt2016flow} \hl{(see Appendix }\ref{app:HewittComp}). This agreement is not only qualitative: the plateau level in our model coincides with the analytical plateau value reported by Hewitt \textit{et al.} \cite{hewitt2016flow} to within a relative error of less than $0.1\%$, before the curves depart from it again for large values of pressure (see Appendix \ref{app:HewittComp} and figure \ref{fig:PlatComp}). This provides a strong validation of the model in the limit where the wall is effectively rigid. As predicted by the asymptotic analysis in section \ref{sec:Asymp}, divergence from the plateau values occurs once $\Pcal$ becomes larger than the product of membrane stiffness $\Scal$ and initial porosity $\Phi$, allowing the flow rate to increase again. This marks the point at which radial expansion of the membrane starts to relieve axial compaction, so that additional pressure can once more drive additional throughput. The inset to Fig.~\ref{fig:Mainresults2}(a) shows representative radial displacement profiles $\chi(z)$ at three pressures along one of these curves. As $\Pcal$ increases, the membrane bulges outward over the central portion of the cylinder, indicating that this renewed growth of $Q_0$ is associated with wall inflation rather than further vertical collapse of the porous skeleton.

Figures~\ref{fig:Mainresults2}(b,c) show that the onsets and scalings of these regimes collapse when expressed in terms of simple parameter combinations. In Fig.~\ref{fig:Mainresults2}(b), we replot the curves from Fig.~\ref{fig:Mainresults2}(a) against $\Pcal/\Scal$. At large $\Pcal$, the branches fall on a common trend, demonstrating that the post-plateau growth of the flow rate is controlled primarily by the ratio of driving pressure to wall stiffness. Figure~\ref{fig:Mainresults2}(c) compares pressure-driven and gravity-driven forcing. The solid curves reproduce the $Q_0(\Pcal)$ data from Fig.~\ref{fig:Mainresults2}(a), while the dashed curves show solutions with $\Pcal=0$ plotted instead against $\Lcal/2$. The near-collapse of these two sets of curves demonstrates that, in this regime, gravity alone is roughly half as effective as imposed pressure at driving flow. This is consistent with the relative scaling for $\Pcal$ and $\Lcal$ obtained by averaging the outer solution for the wall shape from \eqref{eq:F(z)},
\begin{equation}\label{eq:averageraddef}
    \int_0^1 \chi(z) \ \dz = \frac{\Pcal+\Lcal/2}{\Scal+6G(\Phi)},
\end{equation}
so that $\Lcal$ enters with a factor of one-half. Finally, in Fig.~\ref{fig:Mainresults2}(c) we also include solutions computed with a much smaller bending parameter $\Tcal = 10^{-7}$. The weak difference between those curves and the $\Tcal = 10^{-4}$ results shows that, in this part of parameter space, the overall flow--pressure relation is relatively insensitive to the bending stiffness.

\subsection{Soft-wall response without a compaction plateau}

\begin{figure}
	\centering
	\begin{subfigure}{0.48\textwidth}
		\includegraphics[width=\textwidth]{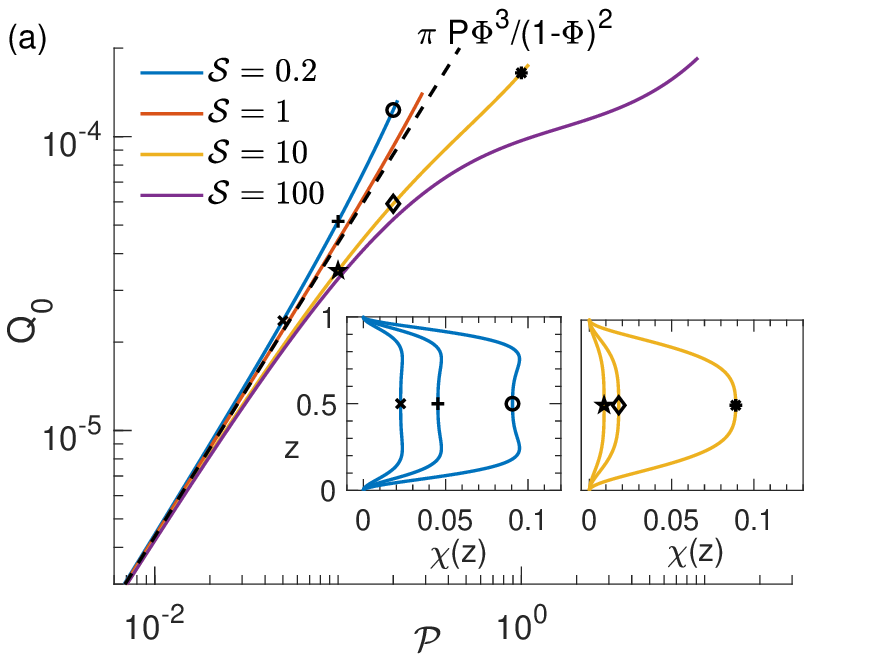}
	\end{subfigure}
	\begin{subfigure}{0.48\textwidth}
	\includegraphics[width=\textwidth]{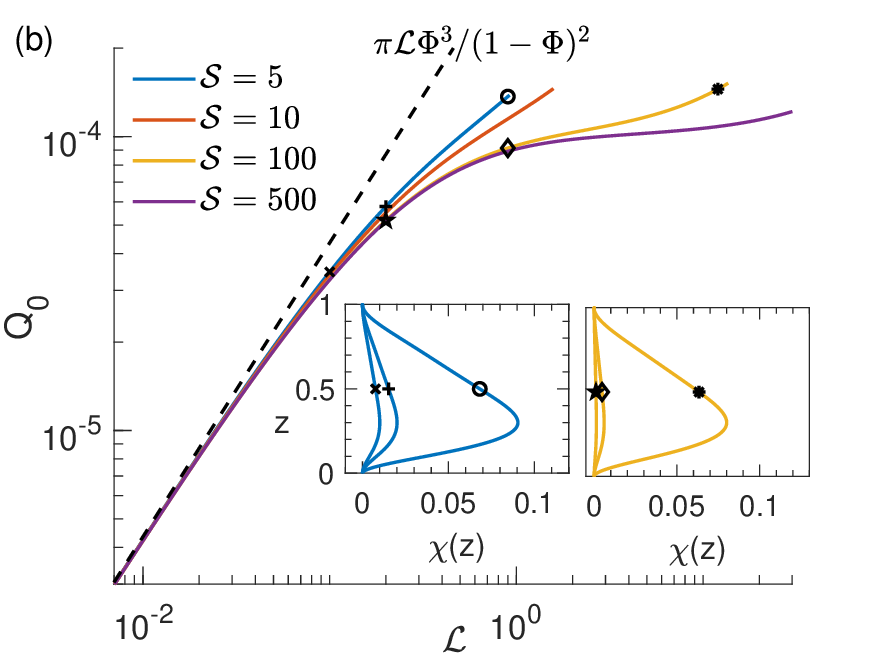}
	\end{subfigure}
	\caption{Flow-rate behaviour for stiffness ratios $\mathcal{S}$ large enough to satisfy the contact condition~\eqref{eq:contcondnond}, but not large enough to produce an intermediate plateau, with undeformed porosity $\Phi=0.05$ and $\mathcal{T}=10^{-4}$. (a) $Q_{0}$ against $\mathcal{P}$ with $\mathcal{L}=0$ for a range of $\mathcal{S}$ indicated in the legends. Inset shows a range of radial profiles taken from the solutions indicated by the shapes. (b) $Q_{0}$ against $\mathcal{L}$ with $\mathcal{P}=0$ for similar values for $\mathcal{S}$, also with a range of profiles shown in the inset.
}
	\label{fig:Mainresults3}
\end{figure}

Figure~\ref{fig:Mainresults3} explores the regime in which the membrane remains stiff enough to make contact with the porous medium, but soft enough that radial expansion becomes important before the medium can develop a fully jammed compaction plateau. Panel~(a) shows the steady flow rate \(Q_{0}\) as a function of imposed pressure \(\mathcal{P}\) at \(\mathcal{L}=0\) for several values of the stiffness ratio \(\mathcal{S}\) (with all other parameters fixed). For small \(\mathcal{P}\) all curves follow the Darcy prediction for an undeformable medium, indicating negligible deformation of both the membrane and the skeleton. As \(\mathcal{P}\) becomes comparable to \(\Phi (\mathcal{S}+2)\) the membrane bulges and deviates from the Darcy prediction. For relatively stiff walls (\(\mathcal{S}\gtrsim 1\)) axial compression of the porous skeleton still dominates, so the flow remains below the undeformable prediction (a sub-Darcy response). As the wall is softened, radial dilation of the column increasingly offsets this compaction, and for the softest admissible wall in panel~(a), \(\mathcal{S}=0.2\), the average porosity rises enough that \(Q_{0}\) exceeds the rigid-medium value, giving a weakly super-Darcy response. For smaller values of \(\mathcal{S}\) the contact condition between the membrane and the porous medium fails in our model, so even softer walls cannot be explored within this unattached-membrane framework. Panel~(b) shows the analogous behaviour for purely gravity-driven flow, plotting \(Q_{0}\) against \(\mathcal{L}\) at \(\mathcal{P}=0\). In this case the contact condition fails at even larger values of \(\mathcal{S}\), and only admissible solutions are shown. The curves again depart from Darcy scaling once $\mathcal{L}$ becomes $O(1)$, but gravity is effectively weaker than the imposed pressure. Even for the softest walls that maintain contact, vertical compaction always outweighs radial dilation. As a result, the response remains at most Darcy or sub-Darcy and no super-Darcy branch appears for gravity alone.

\subsection{Parameter regime map for pressure- and gravity-driven flows}

\begin{figure}[h!]
	\centering
\centering
\includegraphics[width=\textwidth]{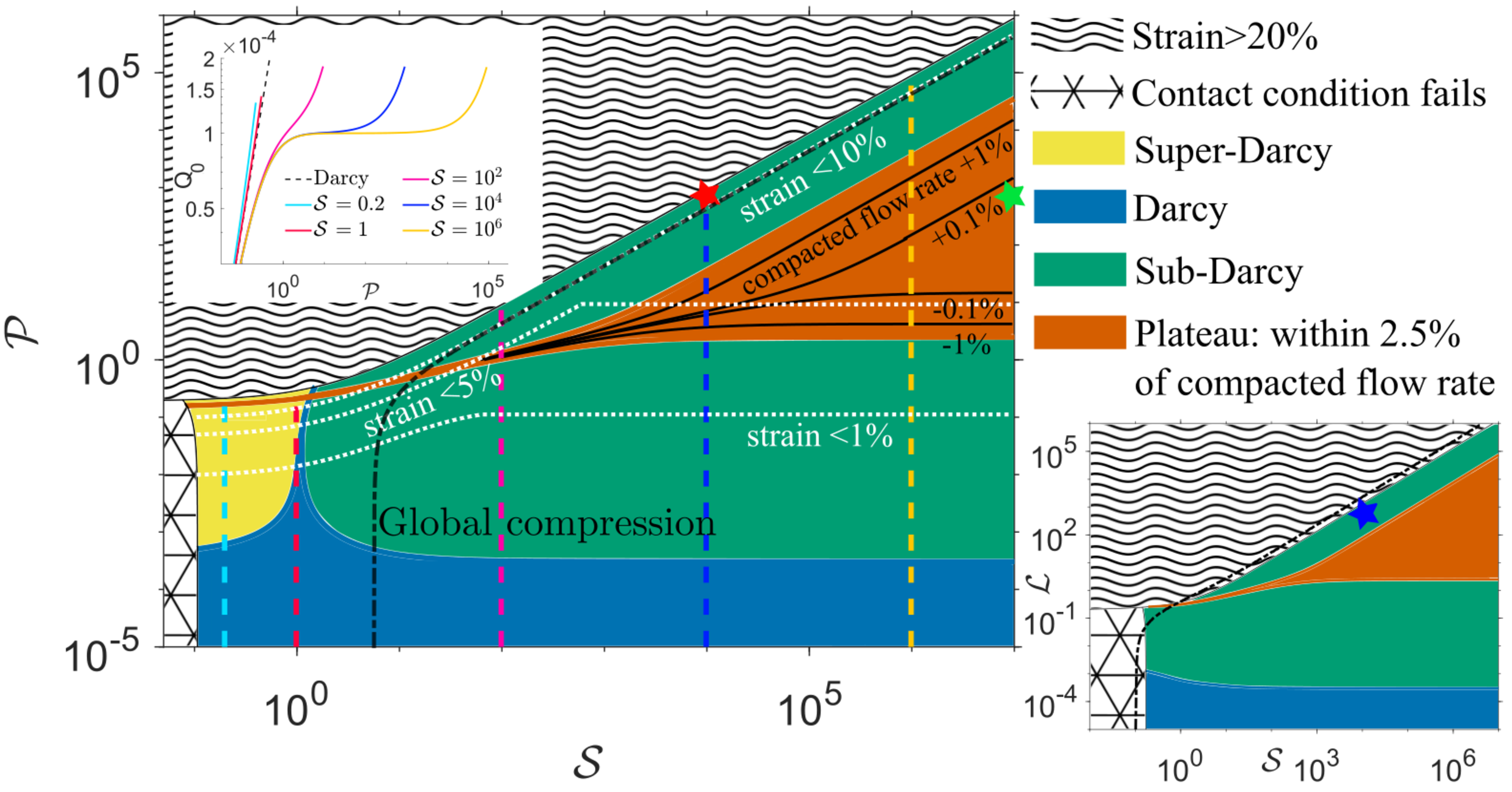}
	\caption{Flow-rate regimes across parameter space for $\mathcal{T}=10^{-4}$ and $\Phi=0.05$. 
Main panel: map of regimes in $(\mathcal{P},\mathcal{S})$ space with $\mathcal{L}=0$. 
Blue region: flow rate within $0.1\%$ of the Darcy prediction $\pi \mathcal{P} \,\Phi^{3}/(1-\Phi)^{2}$ for an undeformable medium of porosity $\Phi$. 
Yellow region (super-Darcy): flow rate exceeds the undeformable prediction. 
Green: flow rate below the undeformable prediction yet still diverging as $\mathcal{P}$ increases. 
Orange: compaction–plateau regime, where the flow rate is approximately constant (within $2.5\%$ of the plateau value). 
The wavy black–white fill marks cases where the maximum strain exceeds $20\%$, which the hatched region shows where the membrane-porous medium contact condition \eqref{eq:contcondnond} fails. 
White dotted contours indicate maximum-strain levels of $1\%$, $5\%$, and $10\%$ as $\mathcal{P}$ increases. Below the black dash–dotted curve, the \hl{solution} satisfies \colorbox{yellow}{$\mathbf{e}_r\cdot \bm{\sigma}^*_{\mathrm{eff}} \cdot \mathbf{e}_r \geq 0$} everywhere (global \hl{compression}); above it, \hl{this condition} is violated somewhere in the cylinder (onset of tension). 
Upper-left inset: selected $Q_{0}$–$\mathcal{P}$ curves for representative $\mathcal{S}$ values, corresponding to the colored dashed transects in the main map. 
Lower-right inset: parameter map in $(\mathcal{L},\mathcal{S})$ space with $\mathcal{P}=0$, showing the absence of a super-Darcy region for purely gravity-driven flow. Stars show the locations of the solutions shown in Fig.\ref{fig:straindef}.}
	\label{fig:Regimes}
\end{figure}

In figure~\ref{fig:Regimes} we summarise these behaviours in a parameter regime map in \((\mathcal{P},\mathcal{S})\)-space at fixed \(\mathcal{L}=0\) and \(\mathcal{T}=10^{-4}\). The colors indicate the qualitative behaviour of the flow rate \(Q_{0}\). The blue Darcy region occupies a band at low \(\mathcal{P}\), corresponding to the weakly forced limit in which both the membrane and skeleton are essentially undeformed. Immediately above this band for $\Scal>1$ lies the green sub-Darcy region, where increasing \(\mathcal{P}\) drives axial compaction and the flow rate lags behind the rigid-medium prediction. At very large stiffness, \(\mathcal{S} \gtrsim 10^{3}\), this feeds into a triangular orange region in which a compaction plateau forms. Changing \(\mathcal{P}\) or $\mathcal{S}$ within this region does not significantly change \(Q_{0}\), as the system has reached the compacted state. However, for all finite $\mathcal{S}$, a value of $\Pcal$ can be reached which unblocks the system and the flow rate starts diverging again. 

The yellow super-Darcy region is confined to softer walls with \(\mathcal{S} \lesssim 1\), and is separated from the green sub-Darcy region by an approximately vertical curve at \(\mathcal{S} \approx 1\) where the effects on the flow rate of axial compaction and radial expansion cancel. The location of this boundary is intuitive: \(\mathcal{S}=O(1)\) marks a balance between the stiffness of the porous skeleton and the stiffness of the membrane. For \(\mathcal{S} > 1\), the \hl{membrane} is relatively stiff and axial compaction dominates over radial expansion, giving sub-Darcy behaviour. For \(\mathcal{S} < 1\), the membrane is more compliant, radial dilatation of the column becomes competitive, and the average porosity can increase enough that the flow becomes super-Darcy. The upper-left inset in figure~\ref{fig:Regimes} shows flow rate behaviours with increasing $\Pcal$, as the coloured flow rate curves correspond to the dashed vertical curves on the main map. At very small \(\mathcal{S}\) the assumptions break down. The hatched region for very small $\mathcal{S}$ marks parameter values for which the contact condition fails.

\hl{The dot--dashed black curve indicates where the radial effective stress becomes tensile, separating solutions in which} \colorbox{yellow}{$\mathbf{e}_r\cdot\bm{\sigma}^*_{\mathrm{eff}}\cdot\mathbf{e}_r\geq0$} \hl{everywhere (on the south--east side of the curve) from solutions in which tensile effective stresses occur somewhere in the column (to the north--west). This distinction is important for granular media such as hydrogel beads, which can sustain compressive but not tensile grain-scale forces. While the other results are calculated using the uncoupled approximation} \eqref{eq:Qz0gen01} and \eqref{eq:nonDFvk3}\hl{, the sign of the radial effective stress near zero is sensitive to the porosity-dependent coupling neglected in} \eqref{eq:nonDFvk3}. \hl{We therefore calculate this boundary using the fully coupled equations} \eqref{eq:Qz0gen01}--\eqref{eq:nonDFvk2}. \hl{For cohesionless granular media, the model is therefore only physically self-consistent in the region south--east of the black dot--dashed line.} The wavy black--white shading superimposed on the map denotes where the axial strain exceeds our nominal small-strain bound, \(|\varepsilon_{zz}|>0.2\). Taken together, the coloured regions, the contact-failure band, and the dot--dashed \hl{tension} boundary provide a compact summary of how membrane stiffness and loading jointly control the transition between Darcy, sub-Darcy, plateau and super-Darcy behaviours, and delineate the subset of parameter space that is relevant for granular media.

To the bottom right of figure~\ref{fig:Regimes}, we show a regime map in $(\Scal,\Lcal)$ space with $\mathcal{P}=0$. The map is very similar to the main figure, however, the values for $\Scal$ at which the contact condition fails are larger, and there is no super-Darcy behaviour for a purely gravity-driven flow. In Appendix \ref{app:smallPhi} figure \ref{fig:RegimeDiagPhi0p01} we show the effects of changing the other parameters $\Tcal$ and $\Phi$. Changing $\Phi$ has minimal impact on the overall structure of the parameter map, but strongly changes strain size which is proportional to $\Phi$. Changing $\Tcal$ also has little effect on the map structure, apart from changing the location of the \colorbox{yellow}{$\mathbf{e}_r\cdot\bm{\sigma}^*_{\mathrm{eff}}\cdot\mathbf{e}_r\geq0$} curve (a smaller $\Tcal$ leads to a smaller valid region for granular media). Coloured stars on figure \ref{fig:Regimes} show the locations in parameter space corresponding to the solutions shown in figure \ref{fig:straindef}.

\subsection{Experimental comparison for rigid and flexible walls}

\begin{figure}
	\centering
	\begin{subfigure}{0.3\textwidth}
		\includegraphics[width=\textwidth]{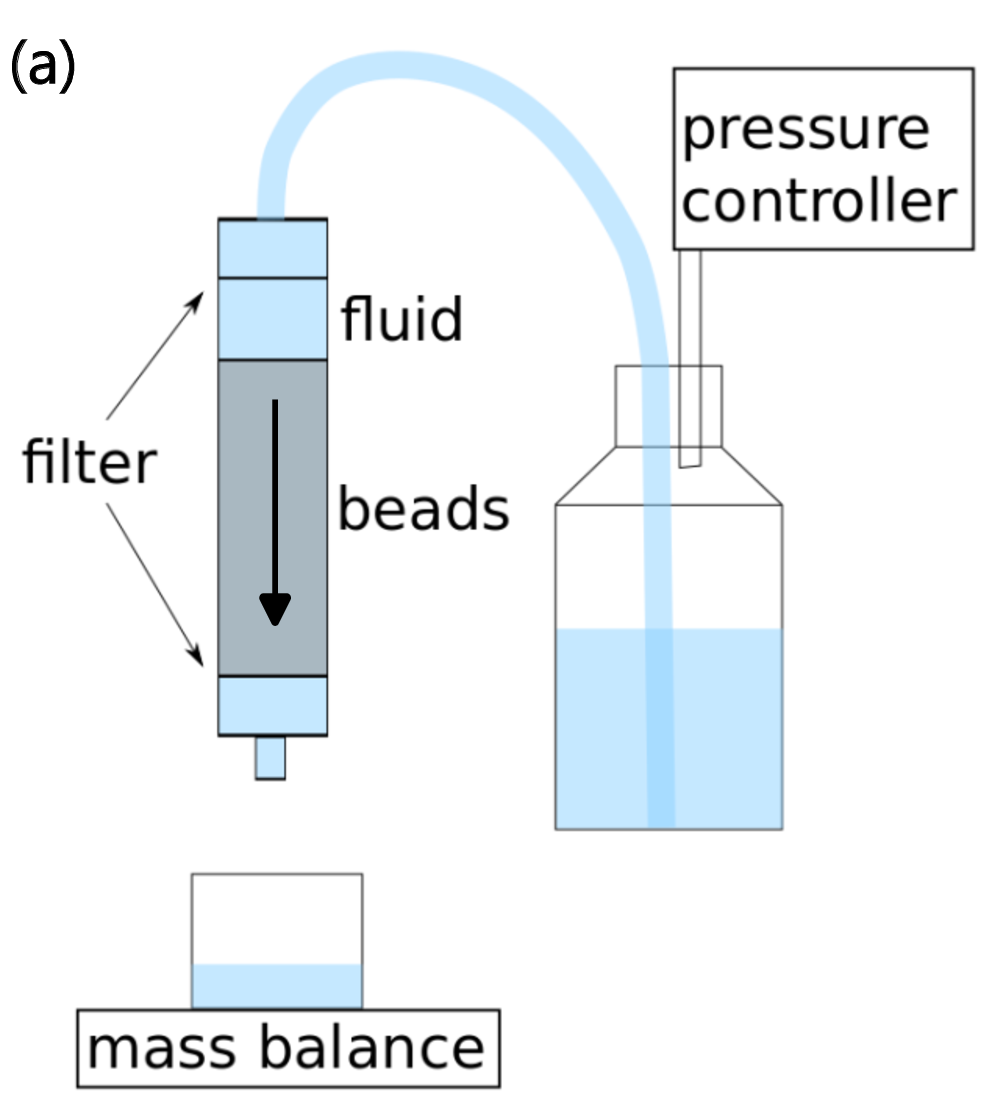}
	\end{subfigure}
\begin{subfigure}{0.67\textwidth}
		\includegraphics[width=\textwidth]{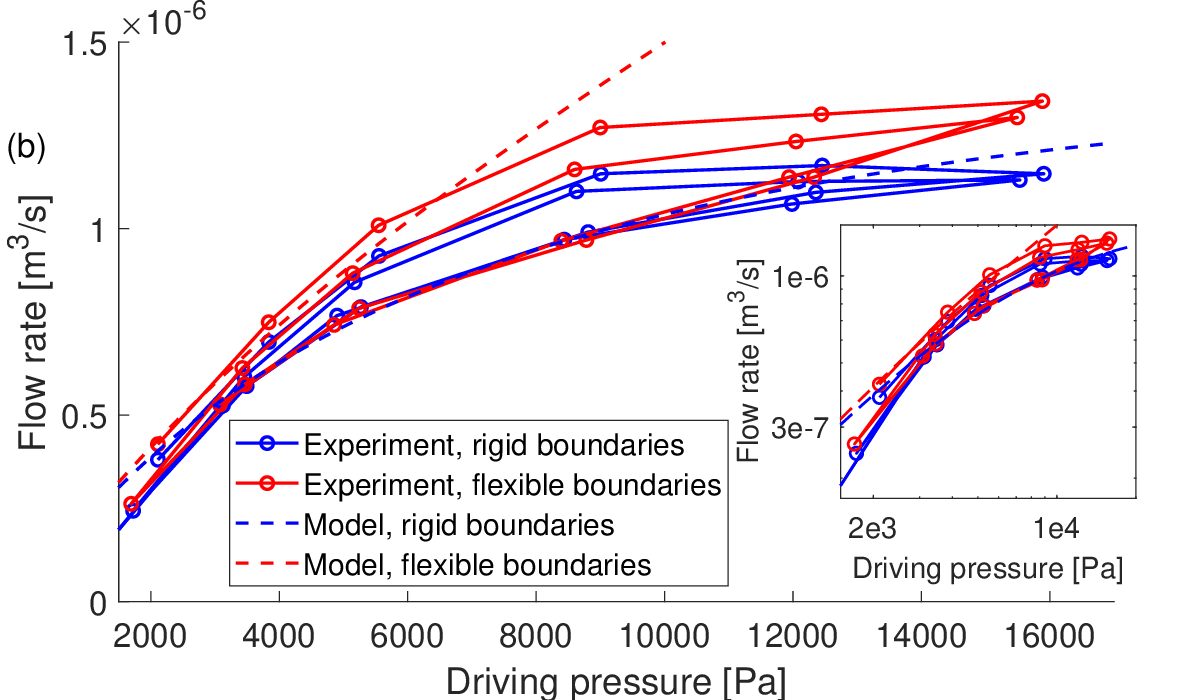}
\end{subfigure}
	\caption{Figure showing experimental setup and results. (a) A schematic of the experiments: fluid flows through a bed of hydrogel beads, first within a rigid cylinder and then within a flexible cylinder; a pressure controller and mass balance capture the pressure and flow rate data. (b) A comparison between the experiments and model results showing qualitative agreement; inset: the same data on a logarithmic scale. \hl{The connected experimental points show the loading and unloading branches of the hysteresis loops.}}
	\label{fig:ExpFig}
\end{figure}

For comparison with the theory, Figure~\ref{fig:ExpFig} shows experimental measurements of the steady flow rate through a column of hydrogel beads first with a stiff radial boundary and then with a flexible radial boundary. The methods used for this experiment are presented in Appendix \ref{sec:expmethods}. \hl{The experimental measurements exhibit hysteresis, with different flow rates observed during increasing and decreasing applied pressure. The experimental points are connected in the order in which they were recorded, to distinguish the loading and unloading branches. Similar hysteresis was observed by Hewitt et al.} \cite{hewitt2016flow} \hl{in experiments, but was not captured by their theoretical model. Likewise, our model describes the steady-state response without accounting for hysteresis.} The blue solid curve corresponds to experiments with a rigid outer wall, for which the deformation is essentially one-dimensional and the appropriate model is the large-\(\mathcal{S}\) limit of Hewitt’s compaction theory. In this case all parameters except the initial porosity \(\Phi\) and the Young’s modulus of the skeleton \(E_s\) are estimated directly from the experiment; \(\Phi\) and \(E_s\) are then fitted using a least squares method to the stiff-wall data, yielding the blue theoretical curve. These fitted values of \(\Phi=0.15\) and \(E_s=2.5\times10^4 \)Pa are then used in the model with an elastic boundary to predict the behaviour in the flexible-wall configuration. The resulting red dashed curve is compared with the flexible-wall experiments (red solid), which show a non-plateauing flow rate. Once the flow departs from the initial Darcy regime, the flow rate continues to increase with driving pressure, consistent with the flexible-boundary regime of the theory in which radial expansion prevents the system from reaching a fully jammed state. Despite some quantitative discrepancies at large forcing, the flexible-wall model captures the overall trend of the experiments and, in particular, the absence of any intermediate plateau.

\section{Discussion}
\label{sec:Disc}

In this study we extended the one-dimensional compaction model of Hewitt et al.\ \cite{hewitt2016flow} to a tall, thin cylindrical porous medium with a laterally deformable, impermeable elastic membrane. Within a small-strain, porosity-dependent poroelastic framework we derived a reduced one-dimensional problem for the axial porosity and membrane shape, and used this to explore how the steady flow rate responds to forcing under combined axial compaction and radial expansion. The rigid-wall limit (formally \(\Scal \to \infty\)) recovers Hewitt's \cite{hewitt2016flow} model. Increasing pressure compacts the outlet region into a boundary layer and the flow rate approaches a genuine compaction plateau, beyond which additional forcing yields essentially no extra throughput. This provides a baseline against which to interpret the effects of finite wall compliance.

Allowing the membrane to be finitely stiff qualitatively reorganises this picture. For sufficiently large but finite stiffness ratios \(S\), the system still exhibits an intermediate plateau that closely matches the rigid-wall behaviour over a broad range of pressures. The outlet region jams, the porosity collapses, and the flow rate becomes nearly pressure-independent. However, this plateau is no longer terminal. Once the imposed pressure becomes comparable to the product of membrane stiffness and initial porosity the compliant wall inflates, relieving outlet compaction and allowing the flow rate to increase again. Thus, for any finite \(\Scal\) the flow--pressure curve ultimately diverges rather than saturating. As \(\Scal\) is reduced, radial expansion becomes important at progressively lower loads. The intermediate plateau shrinks and eventually disappears, leaving purely divergent responses that are sub-Darcy when axial compaction dominates and super-Darcy when wall-driven dilation allows the average porosity to rise above its undeformed value. The regime diagrams built from these solutions summarize how membrane stiffness and forcing control the transition between Darcy, plateau and post-plateau behaviour, and delineate the regions where the model is applicable to granular versus connected porous media.

The experiments with packed hydrogel beads are consistent with the principal theoretical prediction that a flexible boundary eliminates the compaction plateau observed with a rigid boundary. Using only two fitted material parameters inferred from the stiff-wall configuration, the model captures the qualitative trend of the flexible-wall data, including the absence of an intermediate plateau and the monotonic growth of flow rate with pressure. This agreement supports the central mechanistic conclusion that radial compliance prevents the establishment of a fully jammed state.

\hl{For biological context, reported Young moduli of human soft tissues are commonly of order $1$--$100$ kPa, depending on tissue type and measurement conditions} \cite{song2022recent,jawli2024tissue,wells2011medical}\hl{. Reported Young moduli and thicknesses of human amnion and chorion membranes are approximately $47$ MPa and $40$--$50\,\mu$m, and $11$ MPa and $180$--$200\,\mu$m, respectively} \cite{oxlund1990biomechanical}\hl{. Combining illustrative values $E_b^*=20$ MPa, $t_w^*=100\,\mu$m and $E_s^*=10$ kPa gives $\mathcal{S}r_0^*\approx0.2$ m, where $r_0^*$ is a characteristic length scale transverse to a flow. For biological flows with characteristic transverse dimensions much smaller than $0.2$ m, these illustrative material properties imply a $\mathcal{S}\gg1$ regime. However, these properties come from different biological systems, and measured tissue moduli may reflect fluid redistribution and nonlinear solid-network mechanics rather than the solid-phase modulus assumed here. Quantitative application to biological tissues therefore requires system-specific measurements of poroelastic properties, boundary mechanics and physiological pressure gradients. Many biological tissues also have porosities substantially greater than the small values considered in our principal parameter study. Direct measurements of interstitial fluid pressure in solid tumours have also found pressures of several kPa} \cite{milosevic1998interstitial,boucher1990interstitial}\hl{. These measurements suggest that pressure differences comparable to soft-tissue stiffness scales may occur locally, although the resulting dimensionless forcing depends on the effective skeleton modulus and the spatial distribution of the pressure gradient.}

In interpreting our results it is useful to distinguish between disconnected granular packings and connected porous matrices. Our model is formally agnostic to microstructure, but granular assemblies such as hydrogel beads can only support compressive effective stresses, whereas connected skeletons such as sponges or tissues can sustain tension. In the regime map (Fig.~\ref{fig:Regimes}), this implies that parameter combinations that produce tensile effective stresses in the skeleton are more appropriately interpreted as models for connected media. By contrast, the compacted regimes in which \colorbox{yellow}{$\mathbf{e}_r\cdot \bm{\sigma}^*_{\mathrm{eff}} \cdot\mathbf{e}_r\geq0$} everywhere, and the contact condition between membrane and matrix is maintained, are consistent with granular beds. The dot–dashed \hl{tension} boundary and the contact-failure band therefore delimit the portion of parameter space where the model can credibly describe granular systems, while the super-Darcy, \hl{tensional} regimes should be viewed as predictions for connected porous materials.

A natural point of comparison is the recent one-dimensional study of Bouckley et al. \cite{bouckley2025interplay}, who classify soft media by how the flux responds to a pressure drop applied in the direction of gravity. In their framework, with pressure-driven flow aligned with gravity and both ends mechanically constrained, the large-pressure behaviour is controlled entirely by the exponents in the permeability and effective-stiffness laws: “type-1” media have an ever-increasing flux, whereas “type-2” media develop a finite compaction plateau whose existence and height depend on the balance of these exponents. For the constitutive choices adopted here, the rigid-wall limit of our model behaves in an analogous way to their type-2 class. We recover a Hewitt-style plateau in the one-dimensional limit, and our asymptotic solution in negligible gravity plays a similar role to their large-pressure analysis in identifying the effective permeability of a compacted bed. Our work complements Bouckley et al. by showing that once radial compliance is admitted, the presence or absence of a plateau is no longer determined solely by the poroelastic exponents. Any finite wall compliance ultimately destroys the terminal plateau and leads to renewed divergence of the flow rate, with the wall-stiffness ratio $\mathcal{S}$ taking over as the key control parameter. In this sense, their exploration of how varying the permeability and stiffness exponents moves a one-dimensional system between plateau-forming and non-plateau-forming classes is extended here into a two-dimensional setting where an additional degree of freedom reorganizes those regimes and allows for behaviours such as super-Darcy flow that cannot arise in a strictly one-dimensional column with fixed boundaries. We leave it as future work to understand the interplay of $\gamma$ with the different plateau regimes found by Bouckley et al.~\cite{bouckley2025interplay}.

\hl{Our boundary condition on the sidewall is free slip,}
$\mathbf e_r \cdot \boldsymbol{\sigma}^*_{\mathrm{tot}} \cdot \mathbf e_z = 0 $ on $r=1$.
\hl{If, instead, we allow a linear interfacial shear traction, then}
\begin{equation}
\mathbf e_r \cdot \boldsymbol{\sigma}^*_{\mathrm{tot}} \cdot \mathbf e_z
= -\Lambda D_{z0}(z)
\quad \text{on } r=1,
\end{equation}
\hl{where} $\Lambda$ \hl{is a dimensionless interfacial shear parameter
(e.g.} $\Lambda=K_t^*l_0^*/E_s^*$ \hl{for a dimensional interfacial shear stiffness} $K_t^*$, \hl{using the scalings in }\eqref{eq:nondimscheme}).
\hl{This traction persists at steady state and contributes to the cross-sectionally averaged axial force balance through the perimeter-to-area ratio of the cylinder, giving a term of order} $O(\Lambda/\epsilon)$. \hl{Consequently, wall friction is expected to be asymptotically negligible provided} $\Lambda/\epsilon\ll1$, \hl{under which condition the free-slip model remains valid to leading order. For larger interfacial shear stiffnesses, wall friction may alter the deformation and flow-rate behaviour, but quantifying these effects is beyond the scope of the present study.}


A natural concern with any linear–elastic poroelastic model is whether the strains explored are too large for the constitutive assumptions. We emphasize that the undeformed porosity $\Phi$ acts as a practical control parameter for the overall strain level. In Appendix~\ref{app:smallPhi}, Fig.~\ref{fig:RegimeDiagPhi0p01} reproduces the main regime diagram with $\Phi$ reduced from $0.05$ to $0.01$ and shows that the qualitative structure of the map is essentially unchanged, while the strain contours shift upward so that the entire sequence of behaviours discussed in the paper (plateau, its breakthrough,  its disappearance with wall compliance, and the transitions between sub-Darcy and super-Darcy regimes) occurs within an indisputably small-strain range, $\max|\varepsilon_{zz}|\lesssim 0.02$. \hl{Importantly, the nonlinear flow behaviours do not require large elastic strains. When the initial porosity is sufficiently small, even a small deformation can produce a substantial relative change in porosity and hence permeability. Since reducing $\Phi$ preserves the qualitative flow regimes while decreasing the strains at which they occur, we expect the linear-elastic approximation to become increasingly accurate in the small-$\Phi$ limit, provided that the adopted constitutive laws remain valid. Thus, the predicted behaviours are not dependent on the model being applied at large strains, although quantitative predictions at larger $\Phi$ remain subject to finite-strain corrections. Nevertheless, the topology of the regime map and the key onset scalings are expected to remain informative beyond the strict small-strain regime. However, the small initial porosities considered in our main parameter study are below those typical of many of the biological tissues, chromatographic beds and bioreactors motivating this work. While the competition between axial compaction and radial expansion is relevant to these systems, the quantitative predictions and precise regime boundaries obtained at small $\Phi$ should not be extrapolated directly to substantially higher-porosity materials without further investigation.}

\hl{An additional approximation concerns the treatment of gravity in the Lagrangian formulation. As noted in Appendix B, we approximate the gravitational pullback as} $\mathbf g=\mathbf F^T\mathbf G\approx\mathbf G$\hl{, neglecting a deformation-dependent correction first order in strain. Retaining the full gravitational pullback would introduce additional coupling between the axial deformation and gravitational loading. For a prescribed inlet pressure, shortening of the medium would reduce the gravitational contribution to outlet compaction, increasing the outlet porosity, while also modifying the membrane loading and reducing radial expansion. These two effects are expected to have small opposing influences on the flow rate. Alternatively, if the upstream liquid level is maintained at a fixed physical height, the increase in inlet hydrostatic pressure compensates for the reduced gravitational head across the shortened medium, leaving the outlet porosity condition unchanged. In either case, small deformation-dependent corrections to membrane loading remain.}

Beyond the small-strain limitation, several extensions would broaden the range of systems described by the model. Geometric and constitutive extensions include finite aspect ratios, non-axisymmetric and finite-strain deformations, nonlinear or finite-thickness shell mechanics, and a more detailed treatment of transient poroelastic responses. Interfacial effects could be incorporated through tangential wall tethering or friction, wall permeability, loss and recovery of wall–matrix contact, and flow through any resulting gap. For granular beds, further work should account for polydispersity, particle rearrangement and hysteresis, as well as material heterogeneity and anisotropy. At sufficiently large flow rates, inertial and other non-Darcy corrections may also become important.

Taken together, our results highlight that transverse compliance is not a small perturbation to classic one-dimensional compaction, but a qualitative switch that reorganises the flow–pressure relation. In the strict rigid-wall limit the Hewitt–type plateau is genuinely terminal, but any finite wall stiffness ultimately destroys this terminal plateau. Once the driving pressure becomes comparable to the product of membrane stiffness and initial porosity, radial expansion relieves outlet compaction and the flow rate diverges again. The stiffness ratio $\Scal$ therefore acts as a primary design parameter. Large $\Scal$ yields an extended intermediate plateau that mimics the rigid case over several decades in pressure, whereas moderate and small $\Scal$ suppress this plateau and produce purely divergent responses that can be either sub-Darcy or super-Darcy depending on the balance between axial compaction and wall-driven dilation. The regime maps, asymptotic analysis and experimental comparison together provide a compact framework for predicting which of these behaviours will arise in a given system.

\begin{acknowledgments}
R.M. gratefully acknowledges support from the EPSRC Doctoral Prize Fellowship scheme. K.S., A.J. and I.L.C. acknowledge partial support from EPSRC grant [EP/T008725/1]. K.S. and I.L.C. also acknowledge partial support from MRC grant [MR/N011538/1]. R.M. and I.L.C. acknowledge partial support from the Wellcome Leap \textit{In Utero} programme. The authors thank Chris W. MacMinn (University of Oxford) and Oliver E. Jensen (University of Manchester) for helpful discussions.
\end{acknowledgments}

\section*{Data Availability}
All data needed to evaluate the conclusions are present in the paper. The associated computational code is available in the GitHub repository:
\url{https://github.com/RichMcn/Compaction}.

\bibliography{Refs}

\clearpage

\appendix

\section{Tables}
\label{app:tables}

Tables \ref{tab:dimvarsparams} and \ref{tab:keyparamsleadorvars} give the reader a summary of the dimensional and nondimensional variables and parameters used throughout the study.

\begin{table}[h!]
\begin{ruledtabular}
\begin{tabular}{lll}
Symbol & Meaning & Units (SI) \\
\hline
\multicolumn{3}{l}{\textit{Physical quantities}}\\
$r_0^*$    & Initial cylinder radius            & m \\
$l_0^*$    & Initial cylinder length            & m \\
$t_w^*$    & Membrane thickness                 & m \\
$E_s^*$    & Young modulus of skeleton (Poisson ratio of skeleton is assumed to be $1/2$ )      & Pa \\
$E_b^*$    & Young modulus of membrane        & Pa \\
$\nu_b$    & Membrane Poisson ratio             & -- \\
$\mu^*$    & Fluid viscosity                     & Pa\,s \\
$\rho^*$   & Fluid/solid density                 & kg\,m$^{-3}$ \\
$g^*$      & Gravitational acceleration          & m\,s$^{-2}$ \\
$\bar k^*$ & Permeability scale                  & m$^{2}$ \\
$P^*$ & Fluid pressure drop imposed across the cylinder & Pa \\
\multicolumn{3}{l}{\textit{Constitutive functions of porosity}}\\
$k^*(\phi)$ & Permeability & m$^2$
\\
$K^*(\phi)$ & Effective bulk modulus & Pa \\
$G^*(\phi)$ & Effective shear modulus & Pa \\
$M^*(\phi)$ & Effective longitudinal modulus $K^*(\phi)+4G^*(\phi)/3$ & Pa \\
\end{tabular}
\end{ruledtabular}
\caption{A summary of the dimensional quantities used in the model.}
\label{tab:dimvarsparams}
\end{table}

\begin{table}[]
\begin{ruledtabular}
\begin{tabular}{lll}
Symbol & Definition & Role / note \\
\hline
\multicolumn{3}{l}{\textit{Key model parameters}}\\
$\Pcal$  & $P^*/E_s^*$ & Imposed pressure-drop parameter \\
$\Lcal$  & $\rho^* g^*\,l_0^*/E_s^*$ & Gravity/body-force parameter \\
$\Scal$  & $E_b^* t_w/E_s^* $ & Stiffness ratio of membrane vs. solid skeleton  \\
$\Tcal$  & $\epsilon^{4}\, t_w^{2}/\!\bigl(12 (1-\nu_b^2)\bigr)$ & Bending parameter (dimensionless flexural rigidity is $\Tcal\Scal$) \\
$\Phi$ & -- & Reference (undeformed) porosity\\
\multicolumn{3}{l}{\textit{Other parameters}}\\
$\epsilon$ & $r_0^*/l_0^*$ & Initial cylinder aspect ratio (assumed small in all solutions)
\\
$t_w$ & $t_w^*/r_0^*$ & Nondimensional membrane thickness
\\
$\gamma$ & $2\Pcal/(\Scal+2)$ & Pressure–stiffness ratio (used in asymptotic solution) \\
$\Omega$ & $\bigl((\Scal+6G(\Phi))/(4\Scal\Tcal)\bigr)^{1/4}$ & Membrane parameter (boundary-layer thickness is proportional to $1/\Omega$) \\
$\varphi$ & - & Outlet porosity, a function of $\Phi$, $\Pcal$ and $\Lcal$. \\
$B$ & $\varphi/\Phi$ & Ratio of outlet to reference porosity \\
\multicolumn{3}{l}{\textit{Leading order dependent variables}}\\
$\phi_0(z)$ & -- & Porosity field \\
$J(z)$ & $\frac{1-\Phi}{1-\phi_0(z)}$ & Local Jacobian of the deformation \\
$D_{r0}(r,z)$ & -- & Radial displacement  \\
$D_{z0}(z)$ & -- & Axial displacement  \\
$\chi(z)$ & $D_{r0}/r$ & Wall/matrix radial shape;  solves \eqref{eq:nonDFvk2}–\eqref{eq:FVKBoundConds}\\
$p_0(z)$ & -- & Pore pressure profile \\
$U_{z0}$ & -- & Lagrangian Darcy velocity \\
$Q_{0}$ & $\colorbox{yellow}{-}\pi U_{z0}$ & Volumetric flow rate  \\
$u_{z0}(z)$ & $\mathbf{F}_{zz}U_{z0}/J$ & Leading order Eulerian vertical Darcy velocity found via Piola transform
\end{tabular}
\end{ruledtabular}
\caption{Summary of dimensionless dependent variables and parameters. The first block describes the five key parameters upon which our model is dependent. The second block describes other nondimensional parameters used in the model's derivation. The third block describes the leading order dependent variables. Only $D_{r0}$ is shown to be a function of the radial coordinate in the derivation. All of these dependent variables are obtained through the uncoupled solutions for $\chi(z)$ and $\phi_0(z)$.}
\label{tab:keyparamsleadorvars}
\end{table}

\clearpage

\section{Derivation of the pullback of Darcy's law}
\label{app:pullback}

The spatial definition of Darcy's law is
\begin{equation}\label{eq:EulDarcy}
    \mathbf{u}^* = -\frac{k^*(\phi)}{\mu^*}\left(\bm{\nabla}_X^*\,p^* - \rho \mathbf{G} \right) 
\end{equation}
where $\mathbf{u}^*$ is the spatial Darcy velocity field, $\bm{\nabla}_X^*$ is the gradient operator with respect to the Eulerian coordinates and $\mathbf{G}$ is the gravitational vector in the deformed configuration. The Lagrangian Darcy velocity $\mathbf{U}^*$ relates to the Eulerian quantity by the Piola transform for fluxes
\begin{equation}
    \mathbf{U}^* = J \mathbf{F}^{-1}\mathbf{u}^*
\end{equation}
and the pressure transforms using the chain rule while the gravitational contribution is pulled back as part of the spatial driving term,
\begin{equation}
    \bm{\nabla}^* p^* = \mathbf{F}^{*T} \bm{\nabla}^*_Xp^*, \qquad \mathbf{g} =\mathbf{F}^{*T}  \mathbf{G},
\end{equation}
which are relations which can be found in, for example, \cite{aznaran2022transformations} or \cite{holzapfel2002nonlinear}. \hl{Consistent with the small-strain approximation used in the solid mechanics, we neglect the deformation-dependent correction to the gravitational pullback and take} \colorbox{yellow}{$\mathbf g\approx\mathbf G$}, where this neglects a correction first order in the strain. Substituting these into \eqref{eq:EulDarcy} gives us equation \eqref{eq:darcyslawfulleqn}.

\section{Asymptotic simplification of the equations}
\label{app:simp}

The system of equations \eqref{eq:consmass1} to \eqref{eq:mainJacobianeq} under scaling relations \eqref{eq:nondimscheme} reduces to the following system of equations taken at different orders in $\epsilon$
\begin{equation}\label{eq:Continuity2121}
	\frac{\partial (r U_{r0}) }{\partial r}  = 0 
\end{equation}
\begin{equation}\label{eq:Continuity2221}
	\frac{1}{r}	\frac{\partial (r U_{r1}) }{\partial r} + \frac{\partial U_{z0}}{\partial z}
	= 0 
\end{equation}
\begin{multline}\label{eq:Qroeq21}
	\frac{(1+\partial D_{r0}/\partial r)^2U_{r0}+(\partial D_{z0}/\partial r)U_{z0}}{(1-\Phi) k(\phi_0)/(1-\phi_0)}
    \\
    =-\frac{\partial}{\partial r}\left( \left(K(\phi_0)-\frac{2}{3}G(\phi_0)\right)\left( \frac{1}{r}\frac{\partial}{\partial r}(rD_{r0}) +\frac{\partial D_{z0}}{\partial z} \right) +2G(\phi_0)\frac{\partial D_{r0}}{\partial r}\right) 
    \\
	- \frac{\partial }{\partial z}\left( G(\phi_0)  \frac{\partial D_{z0}}{\partial r} \right)
    -2 \frac{\partial}{\partial r}\left( G(\phi_0) \frac{D_{r0}}{r}\right)  ,
\end{multline}
\begin{equation}\label{eq:Dzoeq0121}
	\frac{(1-\phi_0)U_{r0} }{k(\phi_0)(1-\Phi)} \frac{\partial D_{z0}}{\partial r} = \frac{1}{r}\frac{\partial}{\partial r}\left(rG(\phi_0) \frac{\partial D_{z0}}{\partial r}\right) 
\end{equation}
\begin{multline}\label{eq:Qz0eq121}
	\frac{(1+\partial D_{z0}/\partial z)^2U_{z0} + (\partial D_{z1}/\partial r+ \partial D_{r0}/\partial z)U_{r0}}{(1-\Phi)k(\phi_0)/(1-\phi_0)} 
	= 
	- \frac{1}{r}\frac{\partial}{\partial r} \left[ G(\phi_0)r \left(  \frac{\partial D_{r0}}{\partial z} + \frac{\partial D_{z1}}{\partial r} \right) \right]
	\\
	-
	\frac{1}{r}\frac{\partial}{\partial r}\left(rG'(\phi_0)\phi_1 \frac{\partial D_{z0}}{\partial r}\right) 
	- \frac{\partial}{\partial z} \left[ \left( K(\phi_0)-\frac{2}{3}G(\phi_0) \right) \left( \frac{1}{r} \frac{\partial}{\partial r}(rD_{r0}) + \frac{\partial D_{z0}}{\partial z} \right) +2G(\phi_0) \frac{\partial D_{z0}}{\partial z}\right] 
\end{multline}
\begin{equation}\label{eq:Dzoeq0221}
	\frac{\partial}{\partial r} \left(\frac{1}{r}\frac{\partial}{\partial r} \left[  G(\phi_0)r\frac{\partial D_{z0}}{\partial r}  \right]\right)
	=0
\end{equation}
\begin{multline}\label{eq:cauch21}
	-2\frac{\partial^2 }{\partial r\partial z}\left(G(\phi_0) \frac{\partial D_{r0}}{\partial r}\right)
	+ \frac{\partial^2 }{\partial r\partial z}\left(G(\phi_0) \frac{\partial D_{z0}}{\partial z}\right)  
	-2 \frac{\partial^2}{\partial z \partial r} \left(G(\phi_0) \frac{D_{r0}}{r}\right)
	\\ 
	 = - \frac{\partial}{\partial r} \left(\frac{1}{r}\frac{\partial}{\partial r} \left[    G(\phi_0)r\left( \frac{\partial D_{r0}}{\partial z} + \frac{\partial D_{z1}}{\partial r} \right) \right]\right)-	\frac{\partial}{\partial r} \left(\frac{1}{r}\frac{\partial}{\partial r} \left[  G'(\phi_0)\phi_1r\frac{\partial D_{z0}}{\partial r}  \right]\right)
	,
\end{multline}
\begin{equation}\label{eq:porosnondleadingorder21}
	\frac{1}{r}\frac{\partial}{\partial r}(rD_{r0})+ \frac{\partial D_{z0}}{\partial z} = \frac{\phi_0 - \Phi }{1- \Phi},
\end{equation}
\begin{multline}\label{eq:p_r21}
	\frac{\partial p_0}{\partial r} =  \frac{\partial}{\partial r}\left( \left(K(\phi_0)-\frac{2}{3}G(\phi_0)\right)\left( \frac{1}{r}\frac{\partial}{\partial r}(rD_{r0}) +\frac{\partial D_{z0}}{\partial z} \right) 
    +2G(\phi_0)\frac{\partial D_{r0}}{\partial r}\right) 
    \\
    +\frac{\partial}{\partial  r}\left( G(\phi_0)\frac{\partial D_{z0} }{ \partial r}  \right)+2 \frac{\partial}{\partial r}\left(G(\phi_0) \frac{D_{r0}}{r}\right) , 
\end{multline}
\begin{multline}\label{eq:p_z21}
	\frac{\partial p_0}{\partial z}
	=- \mathcal{L}+  \frac{1}{r}\frac{\partial}{\partial r} \left[  G(\phi_0)r\left( \frac{\partial D_{r0}}{\partial z} +\frac{\partial D_{z1}}{\partial r} \right) \right]
	+
	\frac{1}{r}\frac{\partial}{\partial r} \left[G(\phi_0)\phi_1 r\frac{\partial D_{z0}}{\partial r}  \right]  +
	\\
	\frac{\partial}{\partial z} \left[ \left( K(\phi_0)-\frac{2}{3}G(\phi_0) \right) \left( \frac{1}{r} \frac{\partial}{\partial r}(rD_{r0}) + \frac{\partial D_{z0}}{\partial z} \right) +2G(\phi_0) \frac{\partial D_{z0}}{\partial z}\right].
\end{multline}
Equations \eqref{eq:Dzoeq0221} and \eqref{eq:cauch21} were found by taking the divergence of \eqref{eq:cauchysfulleqn}, and equations \eqref{eq:p_r21} and \eqref{eq:p_z21} were found by combining \eqref{eq:cauchysfulleqn} with \eqref{eq:Terzaghi}. Boundary conditions \eqref{eq:velboundcondim} to \eqref{eq:boundconderivDr} reduce to 
\begin{align}
	U_{r0} &=0 \qquad \text{on } r=1 \label{eq:velboundconnondim1}
	\\
	U_{r1} &=0 \qquad \text{on } r=1 \label{eq:velboundconnondim2}
	\\
	p_0 &= \Pcal \qquad \text{on } z=1 \label{eq:presscondnondim1}
	\\
	p_0&=0 \qquad \text{on } z=0 \label{eq:presscondnondim2}
    \\
    \phi_0 &= \Phi \qquad \text{on } z=1
    \label{eq:phicond}
	\\
	\mathcal{T}\mathcal{S} \frac{\dee^4 D_{r0}}{\dz^{4}} + \mathcal{S} D_{r0} &= p_0+ \mathbf{e}_{r}\cdot \bm{\sigma}_{\mathrm{eff}}\cdot \mathbf{e}_{r} \qquad\text{on } r=1 \label{eq:nondimensionalFvK} 
	\\
	\frac{\partial D_{z0}}{\partial r}&=0 \qquad \text{on } r=1
	\\
	\frac{\partial D_{z1}}{\partial r}+\frac{\partial D_{r0}}{\partial z}&=0 \qquad \text{on } r=1 \label{eq:shear2}
    \\
    D_{r0} &= 0, \qquad \text{on } z = 0 ,1 \label{eq:nondimNoRadDef}
    \\
    \frac{\partial D_{r0}}{\partial z} &= 0, \qquad \text{on } z = 0 ,1
	\end{align} 

Equation~\eqref{eq:Continuity2121} shows that \( U_{r0} = 0 \), otherwise $U_{r0}$ would diverge at $r=0$. Equation~\eqref{eq:Dzoeq0121} therefore means that \( D_{z0} = D_{z0}(z) \), and substituting equation~\eqref{eq:porosnondleadingorder21} into the remaining non-zero terms of equation~\eqref{eq:Qroeq21} shows that \( \phi_0 = \phi_0(z) \). Equation~\eqref{eq:porosnondleadingorder21} now shows that $D_{r0}$ must be a linear function of $r$, and this motivates defining the radial displacement profile as \( \chi(z) = D_{r0} / r \). The remaining non-zero terms in equation~\eqref{eq:cauch21} now integrate to
\begin{equation}
\frac{1}{r} \frac{\partial}{\partial r} \left( r \frac{\partial D_{r0}}{\partial z} + r \frac{\partial D_{z1}}{\partial r} \right) = f(z),
\end{equation}
where \( f(z) \) is an arbitrary function of $z$ and must vanish due to boundary condition~\eqref{eq:shear2}. Equation~\eqref{eq:p_r21} now shows that the leading-order pressure \( p_0 \) depends only on \( z \) too. Integrating the remaining nonzero terms in equation~\eqref{eq:p_z21}, and applying \eqref{eq:phicond}, yields the pressure profile
\begin{equation}\label{eq:p_0}
p_0 = \mathcal{P} + \mathcal{L}(1 - z) + M(\phi_0)\left( \frac{\phi_0 - \Phi}{1 - \Phi} \right) - 4G(\phi_0)\chi(z).
\end{equation}
where 
\begin{equation}\label{eq:pwavemod}
    M(\phi_0) = K(\phi_0) +\frac{4}{3}G(\phi_0)
\end{equation}
is the longitudinal modulus, which is the effective resistance to uniaxial compression, combining bulk and shear effects. Enforcing \( p_0 = 0 \) at \( z = 0 \) provides an equation to be solved for the boundary porosity at $z=0$
\begin{equation}\label{eq:varphieqGenapp}
\mathcal{P} + \mathcal{L} + M(\varphi) \left( \frac{\varphi - \Phi}{1 - \Phi} \right) = 0,
\end{equation}
such that \( \phi_0 = \varphi \) at \( z = 0 \).

Now that we have an expression for $p_0$ we can construct the right hand-side of \eqref{eq:nondimensionalFvK}. The \hl{radial} stress is given by
\begin{equation}\label{eq:RADstress}
    \mathbf{e}_{r}\cdot \bm{\sigma}_{\mathrm{eff}} \cdot \mathbf{e}_{r} = - \left(K(\phi_0) - \frac{2}{3}G(\phi_0) \right)\left( \frac{\phi_0-\Phi}{1-\Phi} \right) - 2 G(\phi_0) \chi(z)
\end{equation}
Substituting this and \eqref{eq:p_0} into \eqref{eq:nondimensionalFvK} gives
\begin{equation}\label{eq:nonDFvk2app}
\mathcal{T}\mathcal{S} \frac{d^4 \chi(z)}{dz^4} + \mathcal{S} \chi(z) = \mathcal{P} + \mathcal{L}(1 - z) + 2G(\phi_0)\left( \frac{\phi_0 - \Phi}{1 - \Phi} - 3\chi(z) \right).
\end{equation}
We write \eqref{eq:Continuity2221} as
\begin{equation}\label{eq:Ur1}
	U_{r1}  =- \frac{r}{2}\left(\frac{\partial U_{z0}}{\partial z} \right).
\end{equation}
From boundary condition \eqref{eq:velboundconnondim2}, the bracketed term in \eqref{eq:Ur1} must be zero showing that $U_{z0}$ is a spatial constant. Taking the derivative of \eqref{eq:Qz0eq121} with respect to $z$ we now obtain an uncoupled nonlinear diffusion equation for the leading order porosity
\begin{equation}\label{eq:Qz0gen01app}
	 \frac{\dee}{\dz}\left[  \frac{k(\phi_0)(1+\chi(z))^4(1-\phi_0)}{1-\Phi}
	 \frac{\partial}{\partial z} \left[ M(\phi_0)\left(\frac{\phi_0-\Phi}{1-\Phi} \right) -4G(\phi_0)\chi(z)\right]  \right] = 0 .
\end{equation}

\section{Justification of approximation \eqref{eq:nonDFvk3}}
\label{app:assump}

In this appendix we show that
\begin{equation}\label{eq:assump2}
	\Pcal+\Lcal(1-z) \gg\left|  2G(\phi_0)\frac{\Phi-\phi_{0}}{1-\Phi} -6G'(\Phi)(\Phi-\phi_0)\chi(z)\right| 
\end{equation}
For the choice we make for $G(\phi_0)$ under the Mackenzie model this expression at $z=0$ (where the largest value of the right-hand-side occurs) is
\begin{equation}\label{eq:assump3}
	\Pcal+\Lcal \gg\left|  \frac{2(\Phi-\Phi/(1+9(1-\Phi)(\Pcal+\Lcal)/4))}{3(1-\Phi)} \right| = \left|  \frac{3\Phi(\Pcal+\Lcal)}{2(1+9(1-\Phi)(\Pcal+\Lcal)/4)} \right| 
\end{equation}
which is always $O(\Phi(\Pcal+\Lcal))$.

\begin{figure}
	\centering
\centering
\includegraphics[width=0.7\textwidth]{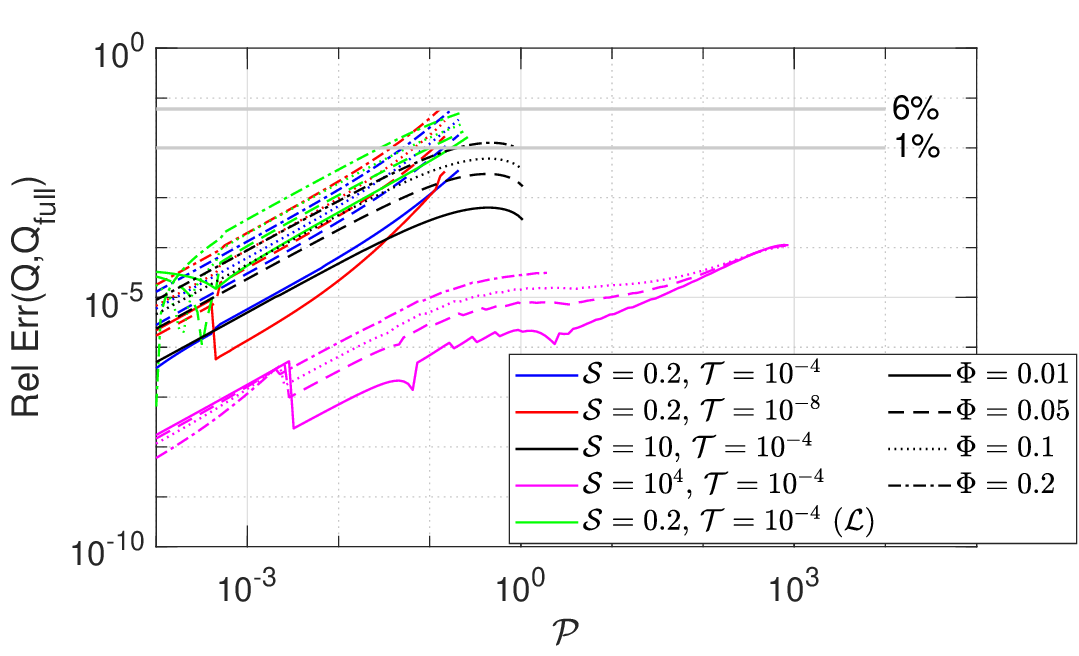}
	\caption{Relative error between solution for flow rate $Q_{0}^f$ calculated from the coupled system (\eqref{eq:NonlinDiff2},\eqref{eq:nonDFvk2}) and the solution $Q_{0}^a$ calculated from solutions of \eqref{eq:NonlinDiff2} with the approximate $\chi$ given by \eqref{eq:F(z)}. Figure shows relative error is negligible for almost all parameter combinations, only rising to $6\%$ for values of $\Scal=10^{-2}$ and larger values of $\Pcal$ where strain approaches our limiting value of $0.2$. Plots are against $\Pcal$ with $\Lcal=0$. Solution is similar against $\Lcal$. }
	\label{fig:FullApproxComp}
\end{figure}

Figure \ref{fig:FullApproxComp} shows the relative error between the flow rates calculated from the full system and the system with the $\chi$ approximation, showing that this error remains small in the whole of the parameter space considered.

\hl{While the approximation} \eqref{eq:nonDFvk3} \hl{accurately captures the flow-rate behaviour, the determination of the sign of individual effective-stress components is more sensitive to the neglected coupling term in} \eqref{eq:nonDFvk2}. \hl{In particular, this term reduces the outward forcing on the membrane during compaction and can influence the radial effective stress near its zero crossing. We therefore evaluate the principal-effective-stress admissibility condition using the fully coupled system} \eqref{eq:Qz0gen01}–\eqref{eq:nonDFvk2}. 

\section{General solution for $\chi(z)$}
\label{app:radialprofilesol}

The equation we consider for the radial deformation is \eqref{eq:nonDFvk3}. We define boundary layer coordinates
\begin{equation}
	\zeta = \frac{z}{\mathcal{T}^{1/4}} \qquad \xi = \frac{1-z}{\mathcal{T}^{1/4}}
\end{equation}
Now look for an expansion
\begin{equation}
	\chi(z,\zeta,\xi) = F_0(z) + f_0(\zeta) +g_0(\xi)
\end{equation}
with boundary conditions $f_0,g_0 \to 0$ as $\zeta,\xi \to \infty$. We get
\begin{equation}
	F_0(z) = \frac{P+L(1-z)}{\mathcal{S}+6G(\Phi)}
\end{equation}
and equations
\begin{equation}
 \frac{\dee^4 f_0}{\dee \zeta^4} + \alpha f_0 = 0
\end{equation}
\begin{equation}
\frac{\dee^4 g_0}{\dee \xi^4} + \alpha g_0 = 0
\end{equation}
where $\alpha = (\mathcal{S}+6G(\Phi))/\mathcal{S}$.
These have solution
\begin{equation}
	f_0 = e^{-(\alpha/4)^{1/4}\zeta}\left(A\cos{((\alpha/4)^{1/4}\zeta)} +B \sin{((\alpha/4)^{1/4}\zeta)} \right)
\end{equation}
\begin{equation}
g_0 = e^{-(\alpha/4)^{1/4}\xi}\left(C\cos{((\alpha/4)^{1/4}\xi)} +D \sin{((\alpha/4)^{1/4}\xi)} \right)
\end{equation}
which gives a matched asymptotic approximation as $\xi,\zeta \to \infty$
\begin{multline}
	\chi(z) = \frac{1}{\Scal \alpha}\left(P+L(1-z)\right) 
    \\
    + e^{-\Omega z}\left(A\cos{(\Omega z)} +B \sin{(\Omega z)} \right) 
	+  e^{-\Omega (1-z)}\left(C\cos{(\Omega(1-z))} +D \sin{(\Omega (1-z))} \right)
\end{multline}
where we have a large parameter
\begin{equation}
	\Omega = \left(\frac{\alpha}{4 \mathcal{T}} \right)^{1/4}
\end{equation}
Applying boundary conditions $\chi=\chi'=0$ \hl{with $e^{-\Omega}\approx 0$} at $z=0$ gives
\begin{equation}
	A =  - \frac{P+L}{\mathcal{S}+6G(\Phi)} \qquad B = A+ \frac{ L }{\Omega (\mathcal{S}+6G(\Phi))}
\end{equation}
Applying boundary conditions $\chi=\chi'=0$ \hl{with $e^{-\Omega}\approx 0$} at $z=1$ gives
\begin{equation}
	C = - \frac{P}{\mathcal{S}+6G(\Phi)} \qquad D = C \colorbox{yellow}{-} \frac{ L}{\Omega (\mathcal{S}+6G(\Phi))}
\end{equation}
which gives equation \eqref{eq:F(z)}.

\section{Mackenzie model of bulk and shear moduli}\label{app:MackenzieModel}

As derived in \cite{mackenzie1950elastic} the Mackenzie model for effective shear and bulk moduli are
\begin{equation}
	\frac{1}{K^*(\phi)} = \frac{1}{(1-\phi)K_s^*} + \frac{3\phi}{4(1-\phi)G_s^*}+O(\phi^3),
\end{equation}
\begin{equation}
	G^*(\phi) = G_s^* - G_s^*\frac{5\phi(3K_s^*+4G_s^*)}{9K_s^*+8G_s^*}+O(\phi^2),
\end{equation}
where $K_s^*$ is the bulk modulus of the solid phase, $G_s^*$ is the shear modulus of the solid phase and $\phi$ is the porosity. In the limit $K_s^* \to \infty$ we obtain 
\begin{equation}
    K^*(\phi) = G_s^*\frac{4(1-\phi)}{3\phi}  \qquad G^*(\phi) = G_s^* 
\end{equation}
to $O(1)$ in $\phi$. As we consider the solid phase to be incompressible, its Poisson ratio is $1/2$ and the Young modulus is equal to three times its shear modulus. We can therefore write the longitudinal modulus and shear modulus as 
\begin{equation}
    M^*(\phi) = E_s^*\frac{4}{9\phi} \qquad G^*(\phi)  = \frac{E_s^*}{3}
\end{equation}

\eqref{eq:BulkandShearModused}.

\section{Implementation of the scheme and calculation of the flow rate}
\label{app:Numerics}

\begin{figure}
	\centering
\begin{subfigure}{0.48\textwidth}
\includegraphics[width=\textwidth]{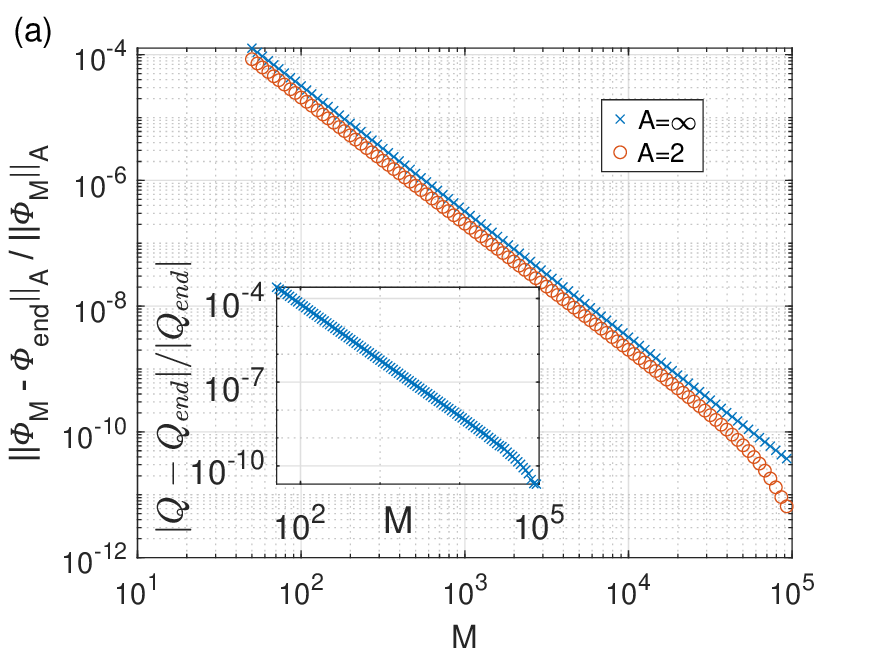}
\end{subfigure}
\begin{subfigure}{0.48\textwidth}
\includegraphics[width=\textwidth]{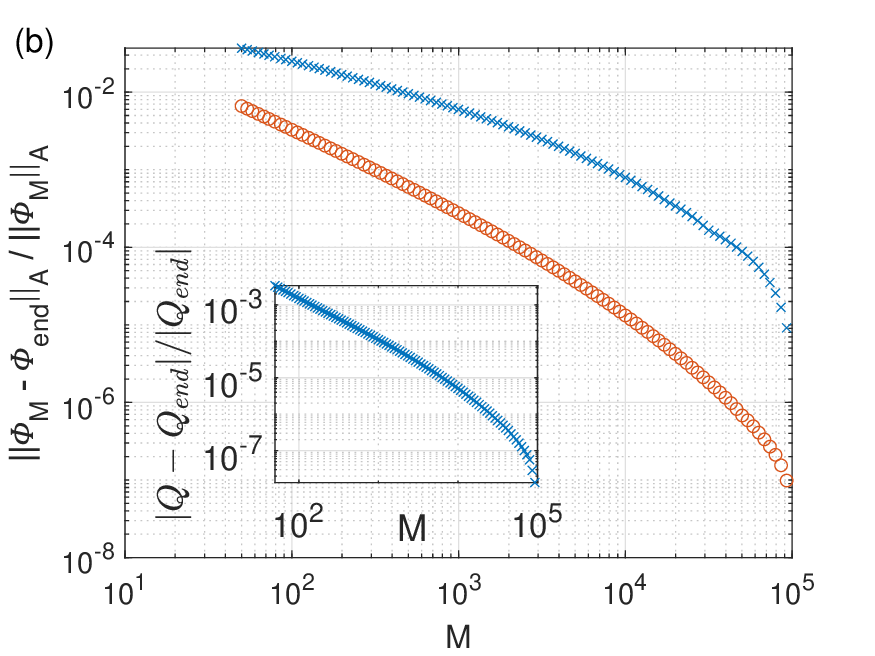}
\end{subfigure}
	\caption{Convergence results for the numerical scheme. (a) Convergence results for $\mathcal{S} = 10$, $\mathcal{T} = 10^{-4}$, $\mathcal{P} = 0.1$, $\mathcal{L} = 0$, and $\Phi = 0.05$. The main plot shows the relative error -- measured by the infinity norm (blue line) and 2-norm (red line) of the relative error between the solution with $M$ grid points and the refined solution $\bm{\phi}_{\text{end}}$ (calculated with $M=10^5$) as a function of $M$. The inset displays the relative error in flow rate between $Q$ and $Q_{end}$ versus $M$. Both indicate second-order convergence with increasing $M$. (b) Convergence for $\mathcal{S} = 10^4$, $\mathcal{T} = 10^{-4}$, $\mathcal{L} = 0$, $\mathcal{P} = 10^2$, and $\Phi = 0.05$. With $\mathcal{P} + \mathcal{L} > 1$, the boundary layer forms, slowing convergence. The main plot shows that the infinity norm error (blue) converges approximately as $M^{-1/2}$, while the two-norm error (red) converges linearly. The inset shows linear convergence of the flow rate $Q$ with $M$.
    }
	\label{fig:conv}
\end{figure}

Equation \eqref{eq:NonlinDiff2} can be rewritten as
	\begin{equation}\label{eq:eqnforSSsol1}
		A(1+\chi(z)) \frac{\dee^2\phi_0}{\dz^2} + A'(1+\chi(z)) \left(\frac{\dee \phi_0}{\dz}\right)^2+(4A-B'(1+\chi))\frac{\dee \chi(z)}{\dz}\frac{\dee \phi_0}{\dz} -B(1+\chi) \frac{\dee^2 \chi(z)}{\dz^2} - 4B\left(\frac{\dee \chi}{\dz}\right)^2 =0,
	\end{equation}
    where
	\begin{equation}
		A = \frac{\Phi\phi_0}{(1-\phi_0)(1-\Phi)}  \qquad A' = \frac{\Phi}{(1-\phi_0)^2(1-\Phi)} \qquad  A'' = \frac{2\Phi}{(1-\phi_0)^3(1-\Phi)}
	\end{equation}
	\begin{equation}
		B = \frac{3\phi_0^3}{(1-\phi_0)} \qquad B' = \frac{3\phi_0^2(3-2\phi_0)}{(1-\phi_0)^2} \qquad B''= \frac{6\phi_0(3-3\phi_0+\phi_0^2)}{(1-\phi_0)^3}
	\end{equation}
 We discretize the domain using a uniform grid with $M$ points
\begin{equation}\label{eq:mesh}
	z=[z_1,z_2 \dots z_M], \quad z_i = (i-1)\Delta z, \quad \Delta z = \frac{1}{M-1}.
\end{equation}
The numerical solution will be approximated using the vector
\begin{equation}
	\bm{\phi} = [v_1, v_2, \dots v_M].
\end{equation}
We approximate \eqref{eq:eqnforSSsol1} at an interior point of the mesh by the residual $\mathbf{q}=[q_1,\dots,q_M]$ defined by the second-order finite difference approximation
\begin{multline}
		q_i= 
		A(v_i)(1+\chi(z_i)) \left( \frac{v_{i+1}-2v_i+v_{i-1}}{\Delta z^2} \right) +A'(v_i)(1+\chi(z_i)) \left(\frac{v_{i+1}-v_{i-1}}{2\Delta z}\right)^2
		\\ +(4A(v_i)-B'(v_i)(1+\chi(z_i)))\frac{\dee \chi(z_i)}{\dz}\left(\frac{v_{i+1}-v_{i-1}}{2\Delta z}\right) - B(v_i)(1+\chi(z_i)) \frac{\dee^2 \chi(z_i)}{\dz^2} -4B(v_i)\left(\frac{\dee \chi(z_i)}{\dz} \right)^2,
\end{multline}
with the components of an interior row of the Jacobian $\mathsf{J}_{ij} = \partial q_j/\partial v_j$ given by
	\begin{multline}
		\mathsf{J}_{i,i-1} = 
		A(v_i)(1+\chi(z_i)) \left( \frac{1}{\Delta z^2} \right) -2A'(v_i)(1+\chi(z_i)) \left(\frac{v_{i+1}-v_{i-1}}{4\Delta z^2}\right) 
        \\
        -(4A(v_i)-B'(v_i)(1+\chi(z_i)))\frac{\dee \chi(z_i)}{\dz} \left(\frac{1}{2\Delta z}\right) ,
	\end{multline}
	\begin{multline}
		\mathsf{J}_{ii}= 	A'(v_i)(1+\chi(z_i)) \left( \frac{v_{i+1}-2v_i+v_{i-1}}{\Delta z^2} \right) -\frac{2}{\Delta z^2} A(v_i)(1+\chi(z_i)) +A''(v_i)(1+\chi(z_i)) \left(\frac{v_{i+1}-v_{i-1}}{2\Delta z}\right)^2
		\\
		+(4A'(v_i)-B''(v_i)(1+\chi(z_i)))\frac{\dee \chi(z_i)}{\dz} \left(\frac{v_{i+1}-v_{i-1}}{2\Delta z}\right) - B'(v_i)(1+\chi(z_i)) \frac{\dee^2 \chi(z_i)}{\dz} -4B'(v_i)\left( \frac{\dee \chi(z_i)}{\dz}\right)^2,
	\end{multline}
	\begin{multline}
		\mathsf{J}_{i,i+1}
		= 		A(v_i)(1+\chi(z_i)) \left( \frac{1}{\Delta z^2} \right) +2A'(v_i)(1+\chi(z_i)) \left(\frac{v_{i+1}-v_{i-1}}{4\Delta z^2}\right) \\
        +(4A(v_i)-B'(v_i)(1+\chi(z_i)))\frac{\dee \chi(z_i)}{\dz}\left(\frac{1}{2\Delta z}\right),
	\end{multline}
with the end components of $\mathbf{q}$ and end rows of $\mathsf{J}$ incorporating the boundary conditions $\phi_1 = \varphi$ and $\phi_M=\Phi$. Next, we initialize the solution vector \( \boldsymbol{\phi}^0 = [v^0_1,v^0_2 , \dots v^0_M] \) using the leading order asymptotic approximation \eqref{eq:leadingorderasympphi} with components
\begin{equation}
	v^0_i = \sqrt{(\Phi^2-\varphi^2)z_i+\varphi^2}.
\end{equation}
The iteration proceeds according to
\begin{equation}
	\boldsymbol{\phi}^{(j)} = \boldsymbol{\phi}^{(j-1)} - J^{-1} \boldsymbol{q}(\boldsymbol{\phi}^{(j-1)}).
\end{equation}
Iteration continues until both of the following convergence criteria are satisfied
\begin{equation}\label{eq:newtonconv}
	\|\boldsymbol{q}\|_{\infty} < \texttt{tol}_\text{it}, \quad \frac{\|\boldsymbol{\phi}^{(j)} - \boldsymbol{\phi}^{(j-1)}\|_{\infty}}{\|\boldsymbol{\phi}^{(j)}\|_{\infty}} < \texttt{tol}_\text{it},
\end{equation}
where we set
\begin{equation}
	\texttt{tol}_\text{it} = 10^{-6}.
\end{equation}
To ensure spatial discretization accuracy, we perform a grid convergence test. We define a sequence of mesh sizes with increasing resolution
\begin{equation}
	M = [100, 200, 400, \ldots],
\end{equation}
doubling the number of grid points at each step. After computing the solution $\boldsymbol{\phi}_{M_i}$ (converged according to \eqref{eq:newtonconv}) on mesh \( M_i \), we interpolate the previous solution from \( M_{i-1} \) onto the finer grid with a cubic method and compare them. The refinement stops when
\begin{equation}
	\frac{\|\boldsymbol{\phi}_{M_i} - \boldsymbol{\phi}_{M_{i-1}}\|_{1}}{\|\boldsymbol{\phi}_{M_i}\|_{1}} < \texttt{tol}_\text{grid},
\end{equation}
with the convergence tolerance set as
\begin{equation}
	\texttt{tol}_\text{grid} =
		10^{-4} 
\end{equation}
This adaptive criterion accounts for the presence of the boundary layer for large \( \mathcal{P} + \mathcal{L} \), which increases the cost of achieving fine-mesh convergence. Finally, the computed solutions have been cross-validated with time-dependent simulations run to steady state using meshes refined near the outlet. Excellent agreement was observed across the parameter space. Convergence of the scheme under grid refinement is shown in figure \ref{fig:conv}.

Flow rate is found by putting the converged solution for $\boldsymbol{\phi}$ into \eqref{eq:flowrate}. The solution is a constant vector apart from small numerical noise. We take the mean of the central half of the vector as our value for the flow rate $Q$. 

We also check our numerics by running the time-dependent method (which has a refined mesh at the boundary layer) to steady-state for many values across the parts of the parameter space where we expect the boundary layer and find excellent agreement.

\section{Derivation of asymptotic solution}
\label{app:asympsol}

We have solution to \eqref{eq:leadordeq1}
\begin{equation}\label{eq:leadingorderasympphi}
	\tilde{\phi}_0 = \sqrt{(1-B^2)z+B^2}
\end{equation}
where $B = \varphi/\Phi$. Defining
	\begin{equation}
		w(z) = \tilde{\phi}_0(z)^3 = \big((1-B^2)z+B^2\big)^{3/2},
	\end{equation}
	then integrating \eqref{eq:leadordeq2} twice gives
	\begin{equation}\label{eq:int-once}
		\tilde{\phi}_0(z)\,\tilde{\phi}_1(z)
		= K_2 + K_1 z 
		+ 3\Omega \int_0^z w(s)\Big(f(s)-f(1-s)\Big)\,ds,
	\end{equation}
	where
	\begin{equation}
		f(u) = e^{-\Omega u}\sin(\Omega u).
	\end{equation}
	Introduce $\lambda=(1-i)\Omega$, so that $f(u)=\Im(e^{-\lambda u})$, where $\Im(.)$ denotes the imaginary part. Then
	\begin{align}
		I_1(z) &= \int_0^z w(s)f(s)\,ds
		= \Im \int_0^z w(s)e^{-\lambda s}\,ds, \\
		I_2(z) &= \int_0^z w(s)f(1-s)\,ds
		= \Im\!\left( e^{-\lambda} \int_0^z w(s)e^{\lambda s}\,ds \right).
	\end{align}
	As $\Omega, |\lambda| \gg 1$, applying $N{+}1$ integrations by parts (Watson’s lemma) yields
	\begin{align}
		\int_0^z w(s)e^{-\lambda s}\,ds
		&= \sum_{k=0}^{N}\frac{w^{(k)}(0)}{\lambda^{k+1}}
		- e^{-\lambda z}\sum_{k=0}^{N}\frac{w^{(k)}(z)}{\lambda^{k+1}}
		+ \frac{1}{\lambda^{N+2}}\int_0^z w^{(N+1)}(s)e^{-\lambda s}\,ds, \\
		\int_0^z w(s)e^{\lambda s}\,ds
		&= e^{\lambda z}\sum_{k=0}^{N}\frac{(-1)^k w^{(k)}(z)}{\lambda^{k+1}}
		- \sum_{k=0}^{N}\frac{(-1)^k w^{(k)}(0)}{\lambda^{k+1}}
		+ \frac{(-1)^{N+1}}{\lambda^{N+2}}
		\int_0^z w^{(N+1)}(s)e^{\lambda s}\,ds.
	\end{align}
    showing each iteration leaving a remainder of size $O(1/|\lambda|)$ the previous iteration.
	Substituting these expansions into \eqref{eq:int-once}, and taking imaginary parts, we obtain
	\begin{multline}
		\tilde{\phi}_0(z)\,\tilde{\phi}_1(z)
		= K_2+K_1 z
		 -\tfrac{3}{2}\,e^{-\Omega z}\,w(z)\,[\sin(\Omega z)+\cos(\Omega z)]
		 \\
		\quad -\tfrac{3}{2}\,e^{-\Omega(1-z)}\,w(z)\,[\sin(\Omega(1-z))+\cos(\Omega(1-z))] + O\!\left(\Omega^{-N-1}\right)
		+ O\!\big(e^{-\Omega}\big).
	\end{multline}
Since $\tilde{\phi}_1(0)=0$ and $\tilde{\phi}_1(1)=0$ we have
\begin{equation}
	 K_2
	= \frac{3}{2} B^3 \qquad K_1
	=\frac{3}{2}(1-B^3) 
\end{equation}
The asymptotic approximation for $\tilde{\phi}$ is
\begin{multline}\label{eq:asympporo}
	\tilde{\phi} = \sqrt{(1-B^2)z+B^2} + 
	\frac{3 \gamma}{2} \Bigg( \frac{B^3+(1-B^3)z}{\sqrt{(1-B^2)z+B^2}}
    \\
    -\,e^{-\Omega z} ((1-B^2)z+B^2)[\sin(\Omega z)+\cos(\Omega z)]
	-\,e^{-\Omega(1-z)} ((1-B^2)z+B^2)[\sin(\Omega(1-z))+\cos(\Omega(1-z))] \Bigg)
\end{multline}
The equation for Lagrangian flow rate, given that we already know the vertical Darcy velocity is a spatial constant, is
\begin{equation}
	\frac{(1-\Phi)Q}{(1-\phi_0)(1+\chi(z))^4} = k(\phi_0) \pi\frac{\dee p}{\dz}.
\end{equation}
which can be solved by dividing by $k(\phi_0)$ and integrating both sides. If we call $I = \int_0^1 (1-\Phi)/((1-\phi_0)k(\phi_0)(1+\chi(z))^4) \dz$, then
\begin{multline}
	\frac{\Phi^3}{1-\Phi} I = \int_0^1 \frac{(1-\Phi(\tilde{\phi}_0+\gamma\tilde{\phi}_1))}{(\tilde{\phi}_0+\gamma \tilde{\phi}_1+\dots)^3(1+\gamma/2(1+\dots))^4} \dz \approx \int_0^1 \left(\frac{1-2\gamma}{\tilde{\phi}_0^3} - 3\gamma \frac{\tilde{\phi}_1}{\tilde{\phi}_0^4} -  \Phi \frac{1}{\tilde{\phi}_0^2}  \right) \dz
\\
	  \approx \frac{2}{B(1+B)} \left( 1-2\gamma -3\gamma \left(\frac{1-B^3}{1-B^2}  \right) + \Phi \frac{B\ln{B^2}}{2(1-B)} \right)
\end{multline}
The next terms are proportional to $1/\Omega$. We need the final cylinder length $l_f$, which is found from 
\begin{equation}
	D_{z0} \approx \int_0^z \left(\frac{\phi_0-\Phi}{1-\Phi} -\gamma \right) \dz = \frac{\Phi}{1-\Phi} \left( \frac{2((1-B^2)z+B^2)^{3/2}}{3(1-B^2)}-z\right) - \gamma z - \frac{2\Phi B^3}{3(1-\Phi)(1-B^2)} 
\end{equation}
\begin{equation}
	l_f = 1 - \gamma + \frac{\Phi}{1-\Phi} \left( \frac{2(1-B^3)}{3(1-B^2)}-1\right) 
\end{equation}
The effective permeability is then
\begin{equation}
	k_{\mathrm{eff}} \approx \frac{\Phi^3B(1+B)}{2(1-\Phi)} \left(  1 + 3\gamma \left( 1+ \frac{1-B^3}{1-B^2} \right) - \frac{\Phi}{1-\Phi} \left( \frac{2(1-B^3)}{3(1-B^2)} -1+(1-\Phi) \frac{B\ln{B^2}}{2(1-B)} \right)  \right)
\end{equation}
Writing $(1-B^3)/(1-B^2) = 1+B^2/(1+B)$, the flow rate is
\begin{equation}\label{eq:asympflowrate}
	Q_{asymp} = \frac{\Phi^3\pi B(1+B)\mathcal{P}}{2(1-\Phi)} \left(  1 + 3\gamma \left( 2+ \frac{B^2}{1+B} \right) - \frac{\Phi}{1-\Phi} \left( \frac{2}{3} \left(1+ \frac{B^2}{1+B} \right)-1+(1-\Phi) \frac{B\ln{B^2}}{2(1-B)} \right)  \right)
\end{equation}

\section{Comparison with the model from Hewitt et al. \cite{hewitt2016flow}}
\label{app:HewittComp}

\begin{figure}
	\centering
\begin{subfigure}{0.75\textwidth}
\includegraphics[width=\textwidth]{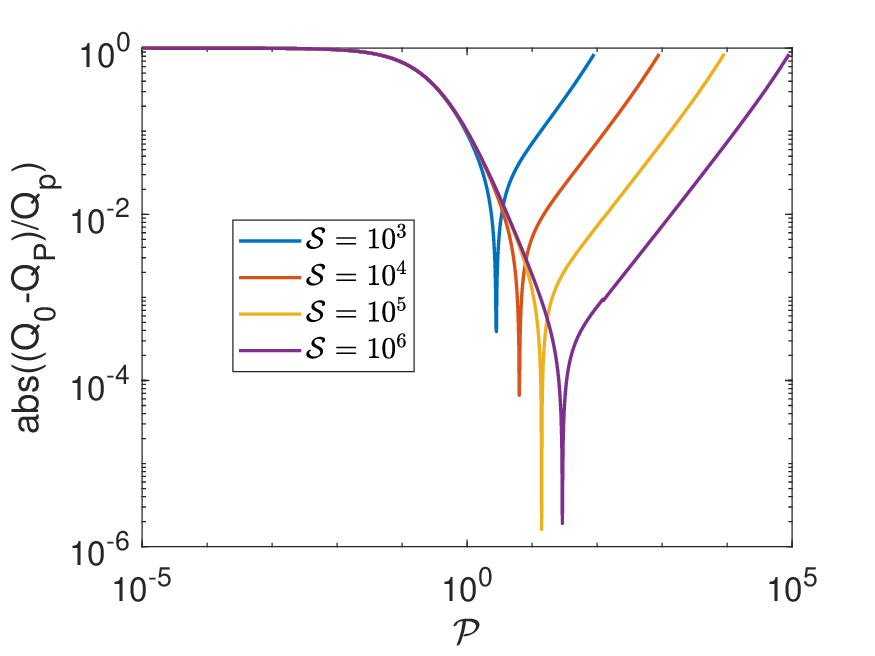}
\end{subfigure}
\caption{Relative error between flow-rate solutions and the analytical expression for the plateau taken from Hewitt. Solutions to \eqref{eq:NonlinDiff2} and \eqref{eq:flowrate} find the analytical expression for the plateau from Hewitt et al. \cite{hewitt2016flow} to within $0.1\%$ before rising again as the breakthrough occurs.}
	\label{fig:PlatComp}
\end{figure}

In figure \ref{fig:PlatComp} we present the relative error between the flow rates $Q_{0}$ found from the solution of \eqref{eq:NonlinDiff2} and \eqref{eq:flowrate} and the analytical solution $Q_p$ presented by Hewitt et al. \cite{hewitt2016flow} with small gravity $\mathcal{L}=0$, bending parameter $\mathcal{T}= 10^{-4}$ and for several values of stiffness ratio $\Scal$ shown. The Hewitt Plateau is
\begin{equation}
Q_p = -\frac{4}{9\Phi} \frac{\pi\Phi^2}{(1-\Phi)^2}\left( \ln{(1-\Phi)}+\Phi \right)
\end{equation}
where the factor of $4/9\Phi$ comes from the different pre-factor we use in \eqref{eq:BulkandShearModused} compared to Hewitt et al. ($4/9$ for the present study and $\Phi$ in \cite{hewitt2016flow}), \hl{along with a factor of $\pi$ to represent the integrated flow rate}. For small values of $\Pcal$ the flow rate is much smaller than the plateau value, but shortly after $\Pcal=1$ is reached, the flow rate hits the transient plateau and the relative error reduces to values much less than $1\%$. For larger values of $\mathcal{S}$, the flow rate stays within a fraction of a percent of Hewitt's plateau for orders of magnitude in $\mathcal{P}$ before the breakthrough occurs and $Q_0$ starts to diverge again. The figure is provided as validation and evidence of consistency with \cite{hewitt2016flow}.

\section{Experimental methods}
\label{sec:expmethods}

To provide context for our theoretical predictions, we conducted a set of simple experiments investigating flow through a porous medium bounded by deformable walls. While limited in scope and not intended as a full validation, these experiments serve to illustrate the qualitative behaviour captured by our model. 

The setup (Figure~\ref{fig:ExpFig}a) comprised a packed bed of hydrogel beads encased in a cylindrical channel with a diameter $20$~mm between two rigid collars which each contained a fine-mesh filter to prevent the drainage of the beads while allowing water to pass. We used sodium polyacrylamide hydrogel beads (JRM Chemical; Beaded Superabsorbents and Snow), which were approximately spherical with diameters of 1–2~mm, when saturated with water. The initial bed height of $\approx 91$ mm. 
Experiments were conducted in two different channels: a modified plastic syringe which was rigid under the pressures applied and a circular silicone sleeve (with wall thickness of 1~mm) connecting the two filter collars. Independent tensile tests on the tubing gave a Young’s modulus of $2.1\times10^6$~Pa. 

For each experiment, flow was driven through the cylindrical channel containing the porous bed by applying a pressure difference between the inlet and the outlet of the channel, which was in contact with atmospheric pressure (see Figure~\ref{fig:ExpFig}a) using a pressure controller (Elveflow). Volumetric flow rates were obtained by weighing the outflow on a mass balance (Kern) as a function of time. The driving pressure was increased \hl{and subsequently decreased} in discrete steps and held for 30~s at each level. The change of the hydrostatic pressure due to changing water levels in the reservoir was taken into account.

To fit the model to experiments we estimated the permeability constant $\bar{k}^*$ by the Kozeny--Carman formula $r_b^{2*}/150$, where $r_b^*$ is the radius of the beads. The least squares error between the experimental data for the rigid case and the model with large $\Scal$ was calculated across a range of values for $E_s$ and $\Phi$, finding a minimum of $E_s = 2.5 \times 10^4$Pa and $\Phi=0.15$. The flow rate for the model for the flexible case was then calculated using these parameters.

\section{Model at a small value of $\Phi$}
\label{app:smallPhi}

\begin{figure}
	\centering
\begin{subfigure}{0.72\textwidth}
\includegraphics[width=\textwidth]{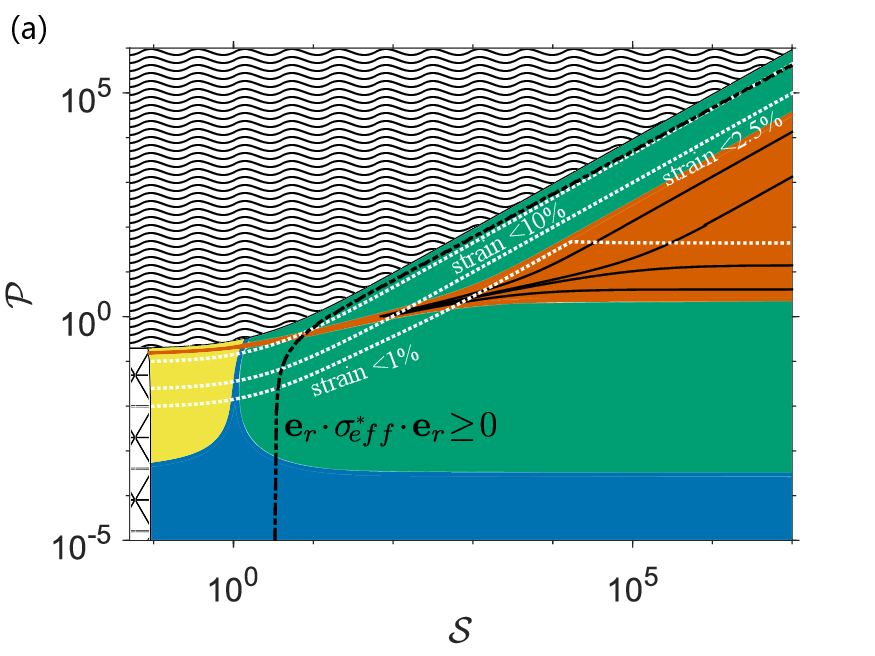}
\end{subfigure}
\begin{subfigure}{0.72\textwidth}
\includegraphics[width=\textwidth]{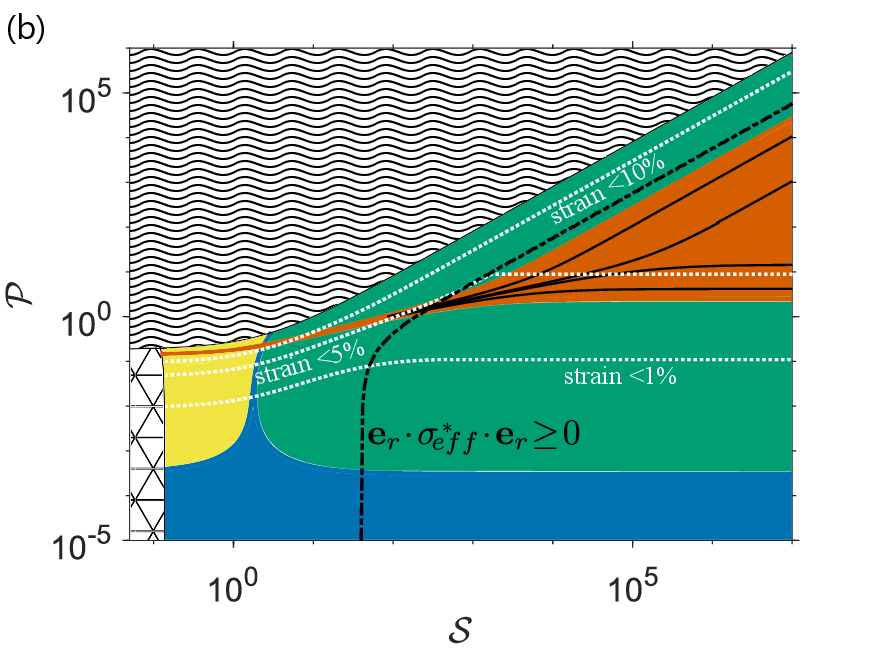}
\end{subfigure}
\caption{Regime diagrams mirroring Fig.~\ref{fig:Regimes}, but with different values of $\mathcal{T}$ and $\Phi$. (a) With an initial porosity of $\Phi=0.01$, rather than $\Phi=0.05$, the white dotted contours of maximum axial strain are shifted upward, while the rest of the diagram remains essentially unchanged. This shows that smaller initial porosity allows the described behaviours to be captured at even smaller strains. (b) The same map as in Fig.~\ref{fig:Regimes}, but with $\mathcal{T}=10^{-7}$ instead of $10^{-4}$. Again, the diagram remains almost unchanged, except that the \colorbox{yellow}{$\mathbf{e}_r\cdot\bm{\sigma}^*_{\mathrm{eff}}\cdot\mathbf{e}_r\geq0$} contour is shifted southeast, reducing the portion of the map that can credibly model disconnected media.}
	\label{fig:RegimeDiagPhi0p01}
\end{figure}

Figure~\ref{fig:RegimeDiagPhi0p01} reproduces the regime map of Fig.~\ref{fig:Regimes} with the undeformed porosity reduced from $\Phi=0.05$ to $\Phi=0.01$. The qualitative structure is essentially unchanged; the principal difference is that the family of strain contours (shown in white) shifts upward in the $(P,S)$ plane, enlarging the low–strain validity wedge. Thus $\Phi$ acts as a convenient control parameter for the overall strain level. For $\Phi=0.01$, all behaviours discussed in the main text—plateau, its disappearance with wall compliance, breakthrough near $P/(S+2)=\mathcal{O}(\Phi)$, and the transitions between sub-Darcy and super-Darcy regimes—occur with $\max|\varepsilon_{zz}|\lesssim 0.02$. Decreasing $\Phi$ further pushes these behaviours to even smaller strains.

\end{document}